\documentclass[twocolumn]{aastex62}

\usepackage{graphicx}	
\usepackage{amsmath}	
\usepackage{amssymb}	
\usepackage{hyperref}
\usepackage{csvsimple}

\hypersetup{
  colorlinks=true,
  urlcolor=blue,
  citecolor= cyan,
  linkcolor = blue
}

\newcommand{\vdag}\

\graphicspath{{./}{figures/}}

\received{---}
\revised{---}
\accepted{---}

\submitjournal{ApJ}

\shorttitle{The phantom halos of the Local Volume in MOND}
\shortauthors{Oria et al.}

\begin{document}

\title{The phantom dark matter halos of the Local Volume in the context of modified Newtonian dynamics}

\author{P.-A. Oria}
\affiliation{Universit\'e de Strasbourg, CNRS, Observatoire astronomique de Strasbourg, UMR 7550, F-67000 Strasbourg, France}

\author{B. Famaey}
\affiliation{Universit\'e de Strasbourg, CNRS, Observatoire astronomique de Strasbourg, UMR 7550, F-67000 Strasbourg, France}

\author{G. F. Thomas}
\affiliation{Instituto de Astrof\'isica de Canarias, E-38205 La Laguna, Tenerife, Spain}
\affiliation{Universidad de La Laguna, Dpto. Astrof\'isica, E-38206 La Laguna, Tenerife, Spain}

\author{R. Ibata}
\affiliation{Universit\'e de Strasbourg, CNRS, Observatoire astronomique de Strasbourg, UMR 7550, F-67000 Strasbourg, France}

\author{J. Freundlich}
\affiliation{Universit\'e de Strasbourg, CNRS, Observatoire astronomique de Strasbourg, UMR 7550, F-67000 Strasbourg, France}

\author{L. Posti}
\affiliation{Universit\'e de Strasbourg, CNRS, Observatoire astronomique de Strasbourg, UMR 7550, F-67000 Strasbourg, France}

\author{M. Korsaga}
\affiliation{Universit\'e de Strasbourg, CNRS, Observatoire astronomique de Strasbourg, UMR 7550, F-67000 Strasbourg, France}
\affiliation{Laboratoire de Physique et de Chimie de l’Environnement, Observatoire d’Astrophysique de l’Universit\'e Joseph Ki-Zerbo (ODAUO), 03 BP 7021,Ouaga 03, Burkina Faso}

\author{G. Monari}
\affiliation{Universit\'e de Strasbourg, CNRS, Observatoire astronomique de Strasbourg, UMR 7550, F-67000 Strasbourg, France}

\author{O. M\"uller}
\affiliation{Universit\'e de Strasbourg, CNRS, Observatoire astronomique de Strasbourg, UMR 7550, F-67000 Strasbourg, France}

\author{N. I. Libeskind}
\affiliation{Leibniz-Institut fur Astrophysik Potsdam (AIP), An der Sternwarte 16, D-14482 Potsdam, Germany}
\affiliation{Univ Lyon, Univ Claude Bernard Lyon 1, CNRS, IP2I Lyon / IN2P3, IMR 5822, F-69622, Villeurbanne, France}

\author{M. S. Pawlowski}
\affiliation{Leibniz-Institut fur Astrophysik Potsdam (AIP), An der Sternwarte 16, D-14482 Potsdam, Germany}

\begin{abstract}
We explore the predictions of Milgromian gravity (MOND) in the Local Universe by considering the distribution of the `phantom' dark matter (PDM) that would source the MOND gravitational field in Newtonian gravity, allowing an easy comparison with the dark matter framework. For this, we specifically deal with the quasi-linear version of MOND (QUMOND). We compute the `stellar-to-(phantom)halo-mass relation' (SHMR), a monotonically increasing power-law resembling the SHMR observationally deduced from spiral galaxy rotation curves in the Newtonian context. We show that the gas-to-(phantom)halo-mass relation is flat. We generate a map of the Local Volume in QUMOND, highlighting the important influence of distant galaxy clusters, in particular Virgo. This allows us to explore the scatter of the SHMR and the average density of PDM around galaxies in the Local Volume, $\Omega_{\rm pdm} \approx 0.1$, below the average cold dark matter density in a $\Lambda$CDM Universe. We provide a model of the Milky Way in its external field in the MOND context, which we compare to an observational estimate of the escape velocity curve. Finally, we highlight the peculiar features related to the external field effect in the form of negative PDM density zones in the outskirts of each galaxy, and test a new analytic formula for computing galaxy rotation curves in the presence of an external field in QUMOND. While we show that the negative PDM density zones would be difficult to detect dynamically, we quantify the weak lensing signal they could produce for lenses at $z \sim 0.3$. 
\end{abstract}


\section{Introduction} 
\label{sec:intro}

The missing mass problem is one of the most pressing questions in present day (astro)physics. The current dominant $\Lambda$ Cold Dark Matter ($\Lambda$CDM) paradigm -- while reproducing with impressive accuracy a large number of observations on all scales -- still has some pending issues, both at cosmological scales \citep[e.g.,][]{2019ApJ...876...85R, 2018Natur.555...67B} and on small scales \citep[e.g.,][]{2017ARA&A..55..343B}. This includes challenges in explaining the diversity of galaxy rotation curve shapes \citep[e.g.][]{2015MNRAS.452.3650O}, the suprisingly low scatter of the Baryonic Tully-Fisher Relation \citep[BTFR,][]{2000ApJ...533L..99M,2016ApJ...816L..14L,2017MNRAS.472L..35D,2019MNRAS.484.3267L}, the high baryon fraction in massive discs \citep[e.g.][]{2019A&A...626A..56P, 2020A&A...640A..70M}, the planes of satellite galaxies problem \citep{2018MPLA...3330004P}, or the prevalence of cold stellar kinematics and the absence of bulges (or massive stellar halos) in most disc galaxies \citep[e.g.,][]{2020MNRAS.498.4386P}. This justifies exploring alternative frameworks, which can range from modifications of the dark matter properties to radical modifications of gravity. 

In galaxies -- especially rotationally-supported ones -- the observed dynamics can be predicted surprisingly well based on the distribution of baryons alone, through Milgrom's law \citep{1983ApJ...270..365M} which is at the heart of the Modified Newtonian Dynamics (MOND) paradigm \citep{1983ApJ...270..365M, 2012LRR....15...10F}. Milgrom's law posits that gravitational accelerations below $a_0 \simeq 10^{-10}{\rm m} \, {\rm s}^{-2}$ approach $(g_N a_0)^{1/2}$, where $g_N$ is the Newtonian gravitational attraction generated by the baryons. In the case where $g \gg a_0$, the dynamics is Newtonian (hence no dark matter-like effect is present), and a smooth transition can be prescribed between the two regimes.

This very simple law directly predicts the observed slope and an effectively zero intrinsic scatter for the BTFR, as well as the universal relation which is observed between the baryonic and dynamical central surface densities of disc galaxies \citep{2016ApJ...827L..19L,2016PhRvL.117n1101M}, and therefore also the diversity of rotation curve shapes, driven by the different surface density of the baryons in different galaxies \citep[e.g.,][]{2019A&A...623A.123G}. The universal relation predicted by MOND is one between the Newtonian gravitational acceleration generated by the baryons and the total one, a relation that is now known as the Radial Acceleration Relation \citep[RAR,][]{2016PhRvL.117t1101M,2017ApJ...836..152L}, which in the $\Lambda$CDM context suggests a strong coupling between DM and baryonic mass that has yet to find a fully satisfying explanation. There have been multiple investigations of the relation in the $\Lambda$CDM context, but while the general shape of the relation can be accounted for indeed \citep{2016MNRAS.456L.127D, 2017ApJ...835L..17K,2017MNRAS.471.1841N,2017PhRvL.118p1103L}, its normalization and small scatter \citep{2017MNRAS.464.4160D}, the latter being actually accounted for solely by observational errors on the inclination and distance of galaxies \citep{2018A&A...615A...3L}, remain puzzling.

With a suitable extension of gravity -- either classical \citep{1984ApJ...286....7B, 2010MNRAS.403..886M} or relativistic \citep[e.g.,][]{2020arXiv200700082S} -- giving rise to Milgrom's law in the weak-field limit\footnote{exactly so in highly symmetric configurations, and approximately in more complex ones}, no dark matter would then actually be needed in galaxies. Moreover, such a framework could avoid the over-formation of bulges in disc galaxies compared to observations \citep{2014A&A...571A..82C, 2016ASSL..418..413C}. In galaxy clusters, the situation is however different \citep[see, e.g.,][]{2008MNRAS.387.1470A,2019A&A...627L...1B}.

This is one of the most radical existing alternatives to the whole {\it Imago Mundi} carried by the $\Lambda$CDM model: it is actually so different that it is sometimes difficult to even phrase the description of stellar systems in the same way. However, despite fundamental differences in the nature of its ingredients, it is useful to ask oneself whether a MONDian Universe really would look so different from our current standard picture. It has for instance been recently demonstrated by \citet{2020arXiv200700082S} that matter power spectra on linear cosmological scales could be very similar within the particular relativistic MOND theory considered by the authors. A lot of work remains to be done in this context, to connect these linear scales to the non-linear regime, and especially to understand whether the MOND ``missing mass" in galaxy clusters could be addressed naturally in such a framework, without resorting to an additional dark matter component (beyond the $k$-essence scalar field playing the role of dark matter on linear scales in that theory). Here, rather than taking a top-down cosmological approach, we will explore how our Local Volume (within $\sim 11$~Mpc) of the Universe looks in the context of MOND. For this endeavour, we resort to the concept of ``Phantom Dark Matter'' (PDM) introduced in, e.g., \citet{1986ApJ...306....9M} in the context of the \citet{1984ApJ...286....7B} version of MOND. The PDM distribution is the distribution of additional dark matter that {\it would} give rise to the MOND gravitational field in Newtonian gravity. This concept allows one to look at a MONDian Universe with Newtonian eyes, which can ease the comparisons with the standard picture. For instance, relations between the stellar and (phantom) halo mass, as well as between the gas and (phantom) halo mass can then be explored in detail. While the PDM distribution can be computed as $\rho_{\rm PDM} = \Delta \Phi/(4 \pi G) - \rho_b$ (where $\rho_b$ is the baryonic density) in any MONDian framework, it takes a central role within the quasi-linear version of MOND \citep[QUMOND][]{2010MNRAS.403..886M} as it is there typically computed {\it before} the MONDian potential itself, as we will detail in Sect.~2. All the investigations of the present paper will be carried out in the QUMOND context, which should qualitatively be generic but could quantitatively deviate from other formulations such as the \citet{1984ApJ...286....7B} version of MOND.

One fundamental aspect in the context of MOND is that the strong equivalence principle must be broken, meaning that the physics in a free-falling frame depends on its gravitational environment. As shown early by, e.g., \citet{1986ApJ...306....9M}, the PDM distribution in the outskirts of a galaxy is therefore heavily influenced by the gravitational environment in which the galaxy resides, and can even be negative in places. This is known as the external field effect (EFE) of MOND: in Newtonian dynamics, the internal dynamics of a system embedded in a constant external field $g_e$ does not depend on that field, but in the MOND paradigm it does, since the total gravitational acceleration (including the external one) is considered. In MOND, Milgrom's law thus only applies when the internal gravitational field is larger than the external one, $g \gg g_e$. If the situation is any different, the EFE has to be taken into account. This is why ultra-diffuse galaxies embedded in an external field can lack a dark matter-like effect in MOND \citep[e.g.,][]{2018MNRAS.480..473F}. This effect has also been used to predict the velocity dispersions in pairs of photometrically indistinguishable dwarf satellites of Andromeda \citep[][where the velocity dispersions of the dwarfs And XIX, And XXI, and And XXV were predicted and verified \citep{2014ApJ...783....7C}]{2013ApJ...766...22M, 2013ApJ...775..139M}. More generally, the gravity from surrounding structures should cause the rotation curves of any galaxy to decline at large radii \citep[e.g.,][]{2016MNRAS.455..449H, 2020ApJ...904...51C,2021arXiv210904745C}, and this decline should depend on the environment. Whether this applies inside galaxy clusters, where the origin of the ``MOND missing mass" remains unclear, is still to be investigated. But in the field, the EFE is an inevitable consequence of the MOND framework. Other possible signatures of the EFE include effects on the secular evolution of discs, including the evolution of bar instabilities \citep{2020ApJ...905..135B}, warp formation \citep{2000ApJ...531L..21B}, or asymmetries in tidal tails from disrupting satellites \citep{2018A&A...609A..44T}.

Maps of Newtonian potential and acceleration up to a distance of 200 Mpc have been established by \citet{2018MNRAS.474.3152D}, in which zones of exceptionally strong external accelerations are highlighted as good testing grounds to probe the EFE. However, high external accelerations can also occur locally. With this in mind, we aim herafter at computing the detailed PDM density distribution around galaxies in our Local Volume, by rigorously taking into account the full non-linearity of MOND. This exercise should both ease the comparison with the standard picture, and help quantify and locate the consequences of the EFE in MOND: zones of negative PDM should actually arise perpendicular to the external field direction \citep[see e.g.][]{2012LRR....15...10F}, and around the point where $g$ and $g_e$ are equivalent. When the escape speed curve of a galaxy can be measured (which is essentially the case only in the Milky Way today), its confrontation with MOND must also take into account the EFE \citep{2007MNRAS.377L..79F, 2008MNRAS.386.2199W, 2018MNRAS.473..419B}, as without it no escape would be possible at all due to a logarithmically divergent potential. 

For such an endeavour, we need, in principle, to numerically solve the generalized Poisson equation of MOND \citep{1984ApJ...286....7B, 2010MNRAS.403..886M}. Numerical solvers have been developed in the past, especially in the context of the development of MOND $N$-body codes \citep{1999ApJ...519..590B, 2007A&A...464..517T, 2008MNRAS.391.1778L, 2009MSAIS..13...89L, 2012MNRAS.421.2598A, 2015MNRAS.446.1060C, 2015CaJPh..93..232L}. Hereafter, for the aforementioned purpose of computing the PDM density distribution around galaxies in the Local Volume, we shall use our own solver, following the approach of, e.g., \citet{2012MNRAS.421.2598A} and \citet{2015CaJPh..93..232L}, using the quasi-linear formulation of MOND \citep[][]{2010MNRAS.403..886M} to directly compute it. In this formulation, the non-linearity of MOND is entirely contained within the computation of the PDM density, and the Newtonian Poisson equation can then be solved to compute the associated gravitational field.

We will first introduce in Section~\ref{sec:methodology} some aspects of MOND and its quasi-linear formulation, and start by exploring some simple scaling relations between the stellar, gas, and phantom halo masses in the case where no EFE is present. In Section~\ref{sec:mapping}, we then make a full and rigorous calculation of the PDM density in the Local Volume, and present a map of the PDM density in the Local Volume, allowing to explore the consequences of the EFE on the scaling relations between the stellar and halo masses. In Section~\ref{sec:MW}, we take advantage of this model of the Local Universe to present an up-to-date fiducial MOND model of the Milky Way (MW) in its environment. Finally, Section~\ref{sec:signatures_of_EFE} illustrates the action of the EFE on galaxies in a strong external field, where we show that the negative PDM density zones could be detected via gravitational lensing.

\section{MOND and phantom halos}
\label{sec:methodology}

\subsection{Quasi-linear MOND}
\label{subsec:QUMOND}

The concept of the theoretical matter that would source the MOND force field in Newtonian gravity, denoted as PDM, is particularly useful for comparisons with various dark matter-based models, and also permits to conceptually visualize some peculiar aspects of MOND such as the external field effect (EFE). In the quasi-linear formulation of MOND \citep[QUMOND][]{2010MNRAS.403..886M}, the whole non-linearity of the framework is contained within the computation of the PDM density, from which one obtains the gravitational potential by solving the Newtonian Poisson equation\footnote{This version of MOND gravity differs slightly from other formulations outside of spherical symmetry \citep{2010PhRvD..81h7304Z}, which means that the EFE can also have a different quantitative effect.}. 

Indeed, in QUMOND, the generalized Poisson equation takes the form \citep{2010MNRAS.403..886M}:
\begin{equation}
\label{eq:QUMOND_poisson}
\Delta \Phi = \nabla\cdot \left[\nu\left(\frac{\left | \nabla\Phi_N \right |}{a_0}\right)\nabla\Phi_N \right]
\end{equation}
where $\Phi$ is the MOND potential, $\Phi_N$ is the Newtonian potential, $a_0$ is the critical acceleration constant mentioned in Section~\ref{sec:intro}, the value of which we fix hereafter at $1.2\times10^{10}$ m$\,$s$^{-2}$ \citep[e.g.,][]{2011A&A...527A..76G}, and $\nu$ is the function allowing for a smooth transition between the Newtonian and MOND regimes. In particular, this $\nu$ function should verify
\begin{equation}
\label{eq:nu_function_1}
\nu(x)\rightarrow 1\ (x\gg 1)
\end{equation}
in order to agree with Newtonian dynamics (if the acceleration is much greater than $a_0$, then Equation~\eqref{eq:QUMOND_poisson} tells us that the MOND potential agrees with the Newtonian one), and 
\begin{equation}
\label{eq:nu_function_2}
\nu(x)\rightarrow 1/\sqrt{x}\ (x\ll 1)
\end{equation}
giving the MOND regime for low accelerations. In this paper we adopt the following $\nu$ function \citep{2005MNRAS.363..603F, 2011A&A...527A..76G, 2012LRR....15...10F, 2018MNRAS.480.2660B}:
\begin{equation}
\label{eq:nu_function}
\nu: x\mapsto \left(\frac{1}{4}+\frac{1}{x}\right)^{1/2}+\frac{1}{2}.
\end{equation}
This is very close to the interpolating function adopted by \citet{2016PhRvL.117t1101M, 2017ApJ...836..152L, 2018A&A...615A...3L} to describe empirically the relation between $\nabla\Phi_N$ and $\nabla\Phi$ in galaxies \citep[see][]{2012LRR....15...10F}. Note however that a modification is needed at the high mass end to pass Solar System constraints \citep{2016MNRAS.455..449H}, but this has no consequence for the study conducted hereafter.

The above Poisson equation can also be recast as
\begin{equation}
\label{eq:MOND_poisson}
\Delta \Phi = 4\pi G(\rho_{\rm PDM}+\rho_b)
\end{equation}
where $G$ is the gravitational constant, $\rho_{\rm PDM}$ is the PDM density, and $\rho_b$ is the baryonic density. Combining Equation~\eqref{eq:QUMOND_poisson} and Equation~\eqref{eq:MOND_poisson}, we readily get a formula for the PDM density:
\begin{equation}
\label{eq:rho_ph}
\rho_{\rm PDM} = \frac{1}{4\pi G}\ \nabla\cdot \left[(\nu\left(\frac{\left | \nabla\Phi_N \right |}{a_0}\right)-1)\nabla\Phi_N \right].
\end{equation}
Numerically, this formula can be discretized and the PDM density can be computed on a grid pattern using finite differences following, e.g., the method used in \citet{2015CaJPh..93..232L}. For the present work, we wrote our own version of this method to compute the phantom density. The value at the $(i,j,k)$ vertex is given by:
\begin{equation}
\begin{aligned}
\label{eq:rho_ph_vertex}
\rho_{\rm PDM}(i,j,k) = &\ \alpha[\Tilde{\nu}\left(\frac{|\nabla\Phi_N(i+1,j,k)|}{a_0}\right)\frac{\partial \Phi_N}{\partial x}(i+1,j,k) \\
& -\Tilde{\nu}\left(\frac{|\nabla\Phi_N(i-1,j,k)|}{a_0}\right)\frac{\partial \Phi_N}{\partial x}(i-1,j,k) \\
& +\Tilde{\nu}\left(\frac{|\nabla\Phi_N(i,j+1,k)|}{a_0}\right)\frac{\partial \Phi_N}{\partial y}(i,j+1,k) \\
& -\Tilde{\nu}\left(\frac{|\nabla\Phi_N(i,j-1,k)|}{a_0}\right)\frac{\partial \Phi_N}{\partial y}(i,j-1,k) \\
& +\Tilde{\nu}\left(\frac{|\nabla\Phi_N(i,j,k+1)|}{a_0}\right)\frac{\partial \Phi_N}{\partial z}(i,j,k+1) \\
& -\Tilde{\nu}\left(\frac{|\nabla\Phi_N(i,j,k-1)|}{a_0}\right)\frac{\partial \Phi_N}{\partial z}(i,j,k-1)]\\
\end{aligned}
\end{equation}
with $\alpha=1/(8\pi Gh)$ where $h$ is the width of a cell, and $\Tilde{\nu}=\nu -1$. Notice that when $\nu=1$, i.e. when in the Newtonian regime, $\tilde{\nu}=0$ and thus there is no PDM. Here the partial derivatives of $\Phi_N$ are also computed with finite differences. The formula for the $x$ direction reads
\begin{equation}
\label{eq:gradient}
\begin{aligned}
\frac{\partial \Phi_N}{\partial x}(i,j,k) &= \frac{1}{12h}[\Phi_N(i-2,j,k)-8\Phi_N((i-1,j,k) \\
&+8\Phi_N(i+1,j,k)-\Phi_N(i+2,j,k)],
\end{aligned}
\end{equation}
with the other directions being treated similarly. 
From Equation~\eqref{eq:rho_ph_vertex}, one can see that the only information needed to compute $\rho_{\rm PDM}$ is thus the Newtonian potential $\Phi_N$. However, it is also clear that the PDM density will depend on the gravitational environment of an object, meaning that a rigorous exploration of the structure of phantom halos around galaxies needs to fully take into account its environment. This is what we will explore in Section~\ref{sec:mapping} for the Local Volume. However, before delving into these detailed calculations which mostly affect the outskirts of the PDM halos around galaxies, it is useful to consider the expected relations between the baryonic contents of a galaxy and its PDM halo in the isolated case.

\subsection{The SHMR and gas-to-halo mass relations in isolated MOND}
\label{subsec:isolated_MOND}

While the distribution of baryons alone (albeit with an influence from baryons located far away) dictates the gravitational field around galaxies in the MOND context, this is not the case in the $\Lambda$CDM context where dark matter is playing the key role. In the latter context, the connection between galaxy and halo properties is of utmost importance, and the most explored bit of this connection is the relation between galaxy stellar mass and halo mass, the so-called stellar-to-halo-mass relation (SHMR). One key feature of $\Lambda$CDM is that the halo mass function and the galaxy stellar mass functions have very different shapes: matching the halo mass function to the observed stellar mass function, known as the ``abundance matching" ansatz, then yields a characteristic SHMR \citep{2004MNRAS.353..189V,2004ApJ...609..482K,2013ApJ...770...57B,2013MNRAS.428.3121M}. This SHMR is non-linear in a log-log plot, and displays a break at around $L_*$ galaxies. While this break is observed when considering early-type galaxies (mostly residing in groups and clusters), this is not the case for spirals in the field, which tend to display a monotonically increasing power-law \citep[e.g.][]{2019A&A...626A..56P, 2019A&A...629A..59P, 2021A&A...649A.119P}.

It is therefore useful to ask what the SHMR would look like in a MOND context \citep[see also][]{2015MNRAS.446..330W}. While a detailed modelling of a large volume around each galaxy is needed to assess in details the effect of the external gravitational field, we can start by considering the consequence of the isolated MOND predictions on the SHMR. In MOND, the fundamental relation is between the total amount of baryons and the gravitational field, hence we first need to consider the observed scaling between the gas and stellar mass in spiral galaxies. For this, we can consider the scaling relation found by \citet{2012ApJ...759..138P} between the HI and stellar mass, with a typical scatter of 0.2~dex
\begin{equation}
\label{eq:stellar_mass_to_H}
{\rm log} \left( \frac{M_H}{M_{\star}} \right)= -0.43 \, \rm{log}(M_{\star})+3.75.
\end{equation}
This, in turn, can be used to compute the total gas mass for each galaxy using a multiplicative factor to account for helium: $M_{\rm gas}=1.4M_H$. This however neglects the molecular gas component which should be small in spiral galaxies.

In the standard context, a pure NFW \citep{1997ApJ...490..493N} profile extending to infinity would have an infinite mass, but the virial mass is usually defined at a radius $r_{200}$ where the mean density is 200 times the critical density of the Universe. Hence in MOND we can do the same and integrate the PDM density profile up to a radius $r_{200}$. To do this we choose in agreement with \citet{2013AJ....145..101K} the value $\rho_{\rm crit} = 1.46\times10^{2}$ M$_\odot$kpc$^{-3}$ for said critical density. In a spherical case (e.g. point mass approximation), the enclosed PDM mass $M_{\rm PDM}(r)$ at a radius $r$ can be computed as
\begin{equation}
\label{eq:isolated_PDM_mass}
M_{\rm PDM}(r) = M_b \times \left(\nu\left(\frac{GM_b}{r^2a_0}\right)-1\right)
\end{equation}
where $M_b$ is the baryonic mass obtained from the stellar mass via Eq~\eqref{eq:stellar_mass_to_H}, and $\nu$ is the interpolating function of Eq~\eqref{eq:nu_function}. For a given stellar mass, applying this at $r_{200}$ yields the isolated MOND $M_{200}$ in terms of PDM.

\begin{figure*}
\centering
  \includegraphics[angle=0,  clip, width=15cm]{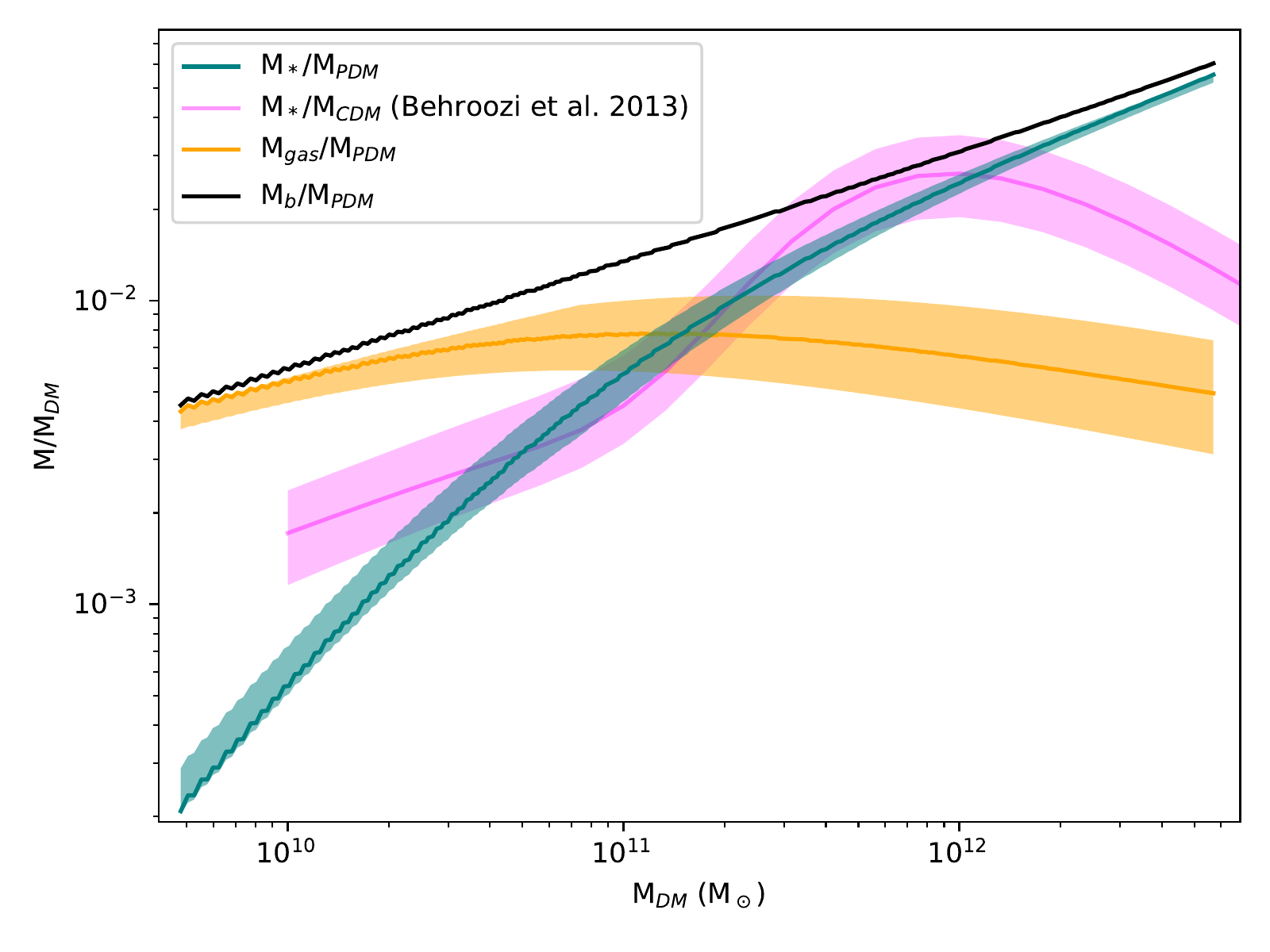}
   \caption{Stellar, gas, and baryonic masses (respectively $M_*$, $M_{\rm gas}$ and $M_b$) over PDM mass $M_{\rm PDM}$ at a computed $r_{200}$ as a function of $M_{\rm PDM}$ in the isolated MOND case. The coloured areas represent the scatter of 0.2 dex in the HI to stellar mass relation of Eq~\eqref{eq:stellar_mass_to_H}. As a means of comparison, the SHMR for $\Lambda$CDM of \citet{2013ApJ...770...57B} (for a redshift $z\sim0.1$) is also plotted.}
\label{fig:isolated_mond}
\end{figure*}

We display the result on Figure~\ref{fig:isolated_mond}, where it appears clearly that the SHMR is monotonically increasing in the isolated MOND case. The SHMR from \citet{2013ApJ...770...57B} (for a redshift $z\sim0.1$) is plotted to highlight the differences with the MOND case. While the values are in good agreement for the range of halo mass considered, the MOND curve shows a very linear behaviour and no break is predicted at high halo masses. Interestingly, if we consider the gas-to-halo mass relation, we find an almost flat relation, meaning that the MOND effect is quite precisely counterbalanced by the observational scaling relation between the stellar and gas mass of galaxies, so as to yield an almost constant $M_{\rm gas}/M_{\rm PDM}$ ratio in the isolated MOND case. We will now explore hereafter how a full treatment of the EFE alters these results. 

\section{Mapping the phantom dark matter in the Local Volume}
\label{sec:mapping}

\subsection{The Updated Nearby Galaxy Catalog (UNGC)}
\label{subsec:UNGC}

To explore numerically the structure of PDM halos in MOND, galaxies cannot be treated in isolation, due to the non-linearity of MOND and the EFE. We will therefore now model the whole Local Volume, and even take into account the effect of large structures outside of it.

Throughout this article, we use a Galactic cartesian coordinate system with the centre of the MW at $(x=0,y=0,z=0)$, with $xy$ being the MW galactic plane. The Local Volume is modeled in our code as a cube of 20 Mpc side length centred on the MW, comprising 800 identical cells in each dimension, giving a resolution of 25 kpc. In order to obtain a map of the PDM distribution for this volume, we require the baryonic gravitational potential. 

We use the UNGC catalog from \citet{2013AJ....145..101K} to get the position and mass (from the $K_s$ band) of 869 galaxies either at a distance less than 11 Mpc from the MW or having a radial velocity with respect to the Local Group inferior to 600 km$\,$s$^{-1}$. 

\begin{figure*}[!htb]
\centering
  \includegraphics[angle=0,  clip, scale = 1]{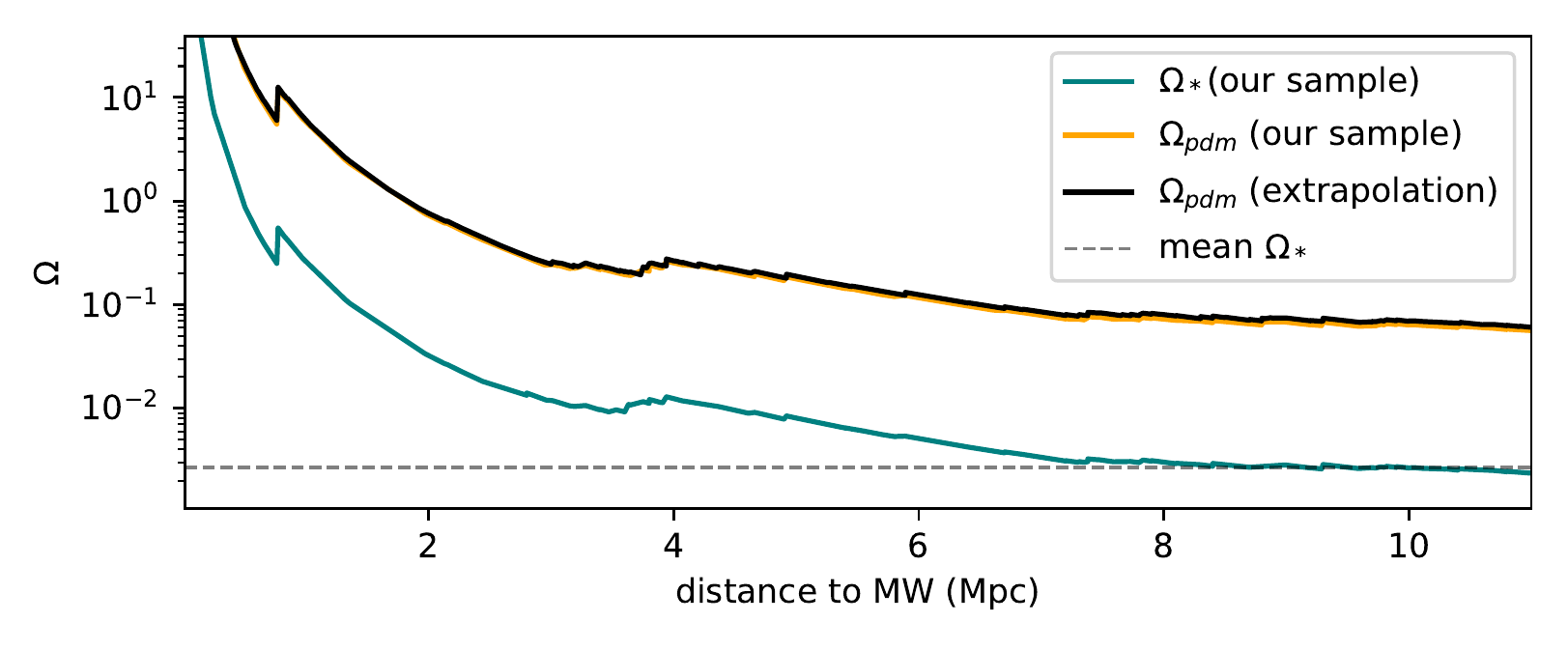}
   \caption{Mean density of matter $\Omega$ as a function of distance to the MW. The dashed line is the mean stellar density of the observable universe computed by \citet{2004ApJ...616..643F}. The teal curve corresponds to the mean stellar density $\Omega_*$ of our sample. The orange curve correspond to the mean PDM density $\Omega_{\rm pdm}$ of our sample. The black curve corresponds to the same value obtained by extrapolating our SHMR of Subsection~\ref{subsec:SHMR} to the whole UNGC. The difference between the orange and black curves is very small, which shows that the galaxies that did not make our cut do not contribute much to the mass of the Local Universe. We notice a particularly good agreement with \citet[Figure 2]{2018AN....339..615K}.}
\label{fig:omegas}
\end{figure*}

More than 50\% of the stellar mass in the Local Volume is contained in the 21 highest luminosity galaxies. Here, for computational purposes, we first choose to only keep galaxies with absolute magnitude $M_{K_s}$ in the $K_s$ band less than or equal to -19. Since we will be interested in computing the `stellar to (phantom) halo mass relation' in the MOND context and in assessing the gravitational environment of the Milky Way in order to produce a fiducial mass model of our Galaxy in MOND, we also include galaxies with apparent magnitude in the $K_s$ band less than or equal to 10 (in order to retain the smaller galaxies in the nearby environment). This gave us a sample of 206 objects, which are listed in the Appendix (Table~\ref{table:local_sources}). The stellar mass $m_{\star}$ in solar masses of a given galaxy was obtained from the following formula:
\begin{equation}
\label{eq:magnitude_to_mass}
M_{\star} = 0.7\times10^{(3.27-M_{K_s})/2.5}
\end{equation}
where 3.27 is the magnitude of the sun in the $K_s$ band \citep{2018ApJS..236...47W}, and we assume a 0.7 the mass-to-light ratio for the $K_s$ band. Our 206 galaxies comprise more than 95\% of the stellar mass of the whole UNGC. We show on Figure~\ref{fig:omegas}, as in \citet{2018AN....339..615K}, the mean stellar density $\Omega_*$ (where the value of the critical density of the Universe is again $\rho_{\rm crit} = 1.46\times10^{2}$ M$_\odot$kpc$^{-3}$) of our sample of galaxies as a function of distance to the MW.

To compute the MONDian PDM, the whole baryonic mass of galaxies needs to be estimated. We again follow \citet{2012ApJ...759..138P} as in Eq~\eqref{eq:stellar_mass_to_H} to estimate the hydrogen mass $m_H$ for each system as a function of their stellar mass, and the gas mass via a multiplicative factor to account for helium: $M_{\rm gas}=1.4M_H$. Note that here we ignore the scatter of this relation, meaning that our results involving stellar and gas masses of individual galaxies will typically underestimate the scatter.

For computational reasons, each of the selected galaxies is added to the Newtonian potential as a point mass with total mass $M_* + M_{\rm gas}$, with the exception of the MW and M31 (and later on, NGC5055). M31 is modeled as an exponential disk based on Equation (2.154) of \citet{2008gady.book.....B} with scale radius $5.9$ kpc, and a Miyamoto-Nagai bulge \citep{1975PASJ...27..533M}, with density parameters adjusted for a total baryonic mass of $1.03\times10^{11}$ M$_\odot$. 

As a first step, the MW is modeled with a Miyamoto-Nagai bulge and disk with parameters from Table 3 of \citet{2013A&A...549A.137I} adjusted for a baryonic mass of $7.5\times10^{10}$ M$_\odot$ of which $6.8\times10^{10}$ M$_\odot$ is in the disk and $7\times10^{9}$ M$_\odot$ is in the bulge. We choose to adopt the above MW set-up for this first computation as it gives an analytical expression for the baryonic gravitational potential. However, we will then upgrade the model to a more realistic configuration in Section~\ref{sec:MW}.

\begin{figure*}[!htb]
\centering
  \includegraphics[angle=0,  clip, scale = 0.44]{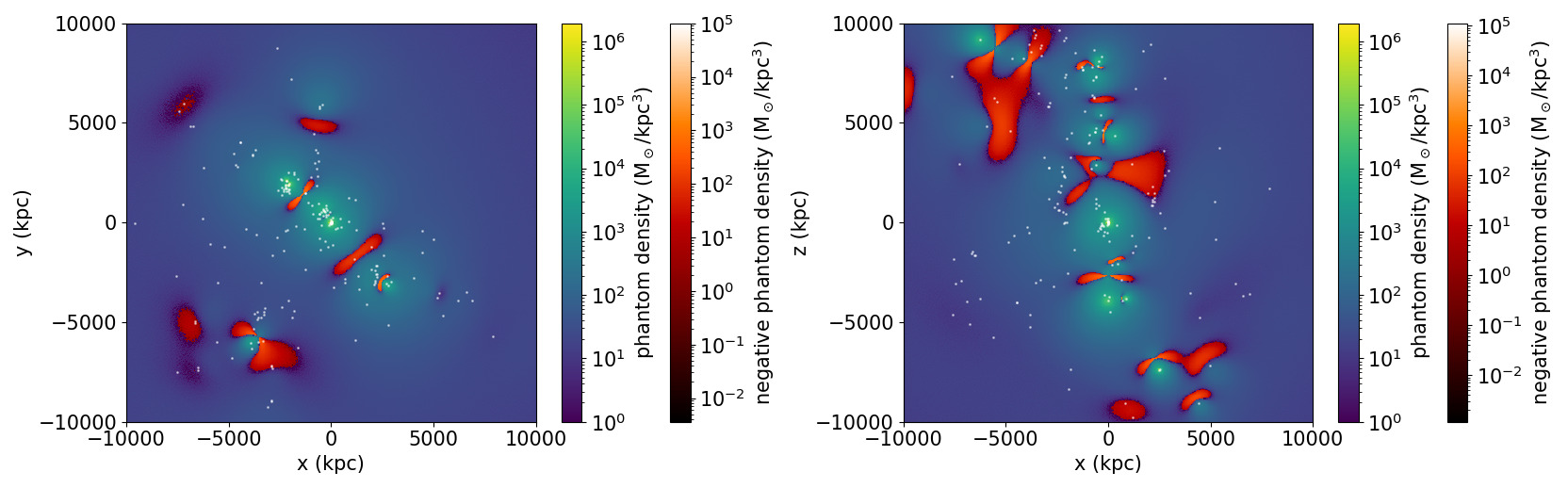}
   \caption{PDM density map of the Local Volume centred on the MW at the position $(x,y,z)=(0,0,0)$ kpc. Plane cuts. \textit{Left panel:} MW galactic plane z=0. \textit{Right panel:} MW edge-on view y=0. White dots are the projections on the respective planes of all the galaxies considered. M31 is at the position $(x,y,z)=(-383,619,286)$ kpc.}
\label{fig:10Mpc_no_ext_sources}
\end{figure*}

\begin{figure*}
\centering
  \includegraphics[angle=0,  clip, width=15cm]{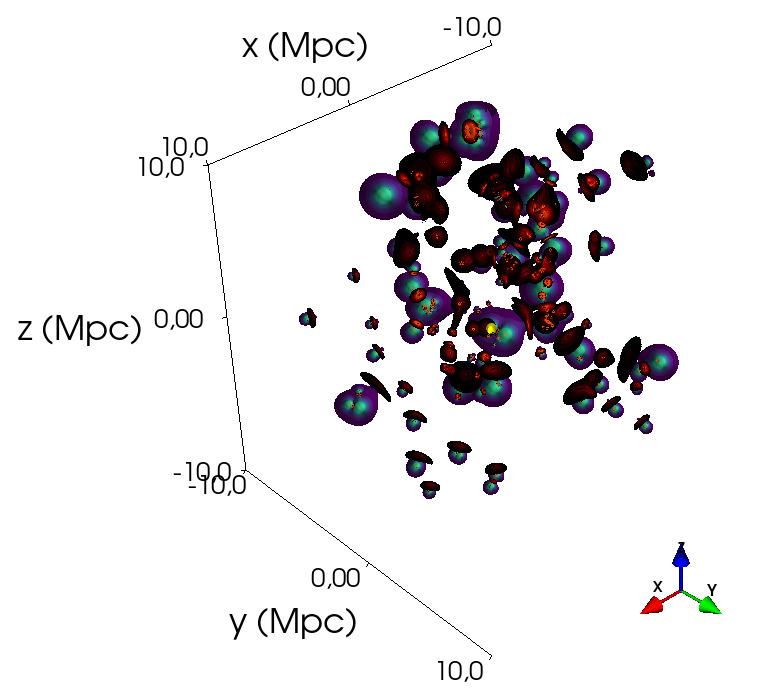}
   \caption{Capture of a 3D contour plot of the PDM density of the Local Volume using only the Local Universe sources of Table~\ref{table:local_sources}. The MW is the yellow ball at the center. The xy plane at $z=0$ is the MW disk plane. Color bars are the same as in Figure~\ref{fig:10Mpc_no_ext_sources}.}
\label{fig:3D_no_ext_sources}
\end{figure*}

\subsection{Phantom dark matter maps}

Equipped with the theoretical tools and catalogue described in the previous sections, we can now use Equation~\eqref{eq:rho_ph_vertex} to compute the PDM density. The result of this first computation is showcased in Figure~\ref{fig:10Mpc_no_ext_sources} for plane cuts of the MW disk plane (z=0) and the edge-on view (y=0). Interestingly, zones of negative PDM density are immediately apparent, mostly perpendicular, for each galaxy, to the direction of the Local Group, which is the dominant EFE source for most of the galaxies sampled. In addition to the plane cuts shown in Figure~\ref{fig:10Mpc_no_ext_sources}, we show in Figure~\ref{fig:3D_no_ext_sources} a capture of a 3D contour plot for the PDM density obtained by using the Mayavi Python package \citep{2011CSE....13b..40R}. The 3D visualization allows us to see the actual shape of the negative PDM zones due to the EFE without projection effects. Furthermore, it allows us to depict many more interactions than would appear on the 2D plane cuts. 

This first result however ignores the influence of larger scales on the Local Volume, which we will now add to highlight this peculiarity of MOND that larger scales can never be ignored when modelling small ones.

\paragraph{Adding galaxy clusters}
The dominant EFE inside the Local Volume is indeed from the influence of sources located outside it, such as galaxy clusters and superclusters. To estimate their influence, we select the most important sources from Cosmicflows-3 \citep{2016AJ....152...50T} in terms of Newtonian gravitational field at the MW. Those sources located outside the Local Volume are included as point masses to the computation. The chosen sample can be found in Table~\ref{table:external_sources}. It has long been known that in the MOND context, the dynamics of galaxy clusters cannot be solely explained by their baryonic content, and an extra source of mass is required \citep[e.g.,][]{1999ApJ...512L..23S, 2008MNRAS.387.1470A}. Various hypotheses have been proposed for this residual missing mass, including hot dark matter \citep{2010MNRAS.402..395A, 2020MNRAS.499.2845H}, baryonic dark matter in the form of cold dense molecular clouds \citep{2008NewAR..51..906M}, and massive gravitating fields that give rise to MOND on small-scales but could behave as DM both on the scale of the CMB and on cluster scales \citep{2020arXiv200700082S}. Here we thus estimate a MOND dynamical mass $M_{\rm MOND}$ for these clusters by assuming that at the virial radius, the $\Lambda$CDM and MOND acceleration should coincide. In our point mass approximation, this leads to
\begin{equation}
\label{eq:MOND_mass}
M_{\rm MOND} = \frac{G M_{\rm CDM}^2}{r_{200}^2 a_0}
\end{equation}
where $G$ is the gravitational constant and $r_{200}$ is used as a substitute to the virial radius, and computed via\footnote{This is obtained by equalling $200\rho_{\rm crit} = 200\times3H_0^2/(8\pi G)$ to $\rho_{200} = 3 M_{200}/(4 \pi r_{200}^3$).}
\begin{equation}
\label{eq:r_200_cluster}
r_{200} = \left(\frac{G M_{\rm CDM}}{100 H_0^2}\right)^{1/3}
\end{equation}
where $H_0$ is the Hubble constant. Table~\ref{table:external_sources} then gives the Newtonian acceleration $g_N$ generated by $M_{\rm MOND}$ of each cluster at their estimated $r_{200}$. Note however that this radius in principle does not play any specific role in the MOND context. Therefore, we checked that our procedure yields a reasonable lower bound on $M_{\rm MOND}$ by converting full NFW enclosed mass profiles to $M_{\rm MOND}(r)$ profiles for the range of virial masses corresponding to the clusters considered here. These $M_{\rm MOND}(r)$ profiles typically reach a maximum value of $M_{\rm MOND}$, which we find to be systematically only $5\% - 15\%$ higher than our estimate of $M_{\rm MOND}$ at $r_{200}$. Our values of $M_{\rm MOND}$ estimated at $r_{200}$ are therefore a reasonable (lower bound) estimate, and the high values we obtain show that it is very reasonable to assume that their enclosed mass should not be significantly altered itself by an EFE from large-scale structure. We also give in Table~\ref{table:external_sources} the Newtonian acceleration $g_N$ generated by each of these sources at the position of the MW. One can see that some clusters generate a Newtonian acceleration of several times $10^{-4}a_0$, whilst it had been estimated in \citet{2007MNRAS.377L..79F} that the external field caused by the Great Attractor on the MW is of the order of $0.01 a_0$, i.e. that the Newtonian gravitational field $g_N$ of the Great Attractor at the position of the MW is of the order of $10^{-4}a_0$. Furthermore, we computed the Newtonian acceleration at the MW for all the sources in the MCXC \citep{2011A&A...534A.109P} catalog: while we cannot include the whole catalog for computational reasons, we found that all the top contributors (the Virgo, Perseus, Centaurus and Coma clusters) are already in our sample picked from Cosmicflows-3, and that the first source absent from our sample has a Newtonian gravitational field one order of magnitude below that of Virgo.

\begin{figure*}[!htb]
\centering
  \includegraphics[angle=0,  clip, scale = 0.44]{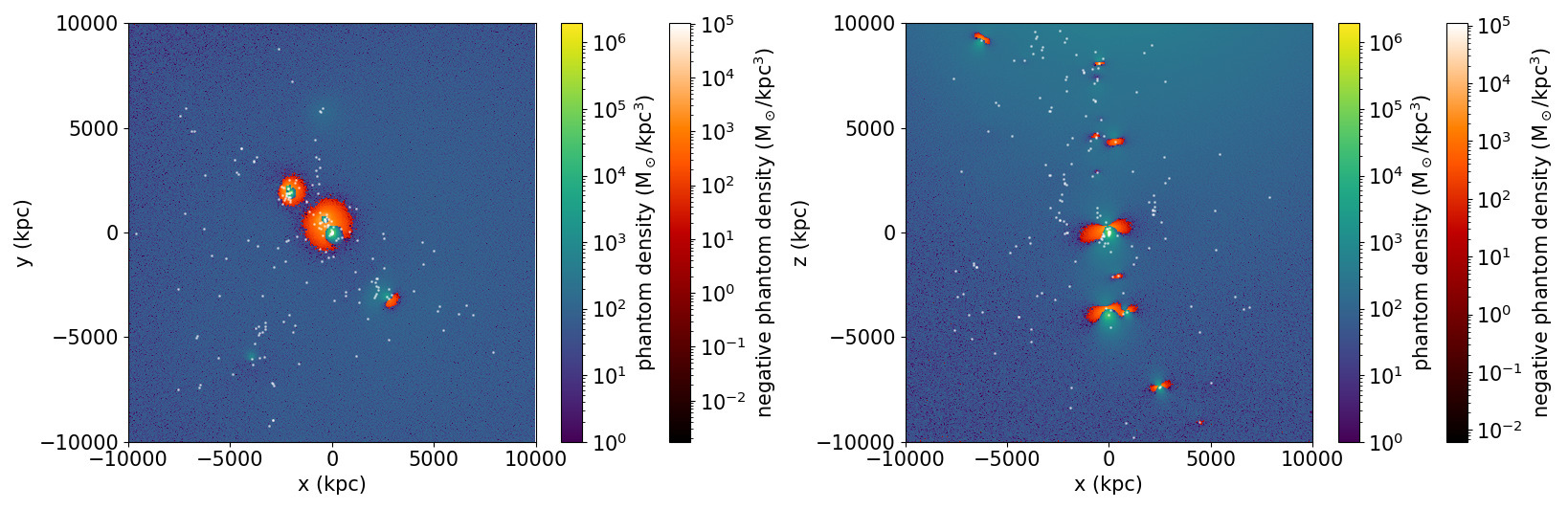}
   \caption{PDM density map of the Local Volume centred on the MW at the position $(x,y,z)=(0,0,0)$ kpc when massive distant sources are added. Plane cuts. \textit{Left panel:} MW galactic plane z=0. \textit{Right panel:} MW edge-on view y=0. White dots are the projections on the respective planes of all the galaxies considered. M31 is at the position $(x,y,z)=(-383,619,286)$ kpc. The Virgo supercluster, source of the dominating EFE in the Local Universe, is at the position $(x,y,z)=(0.38,2.26,17.60)$ Mpc.}
\label{fig:10Mpc_everything}
\end{figure*}

\begin{figure*}
\centering
  \includegraphics[angle=0,  clip, width=15cm]{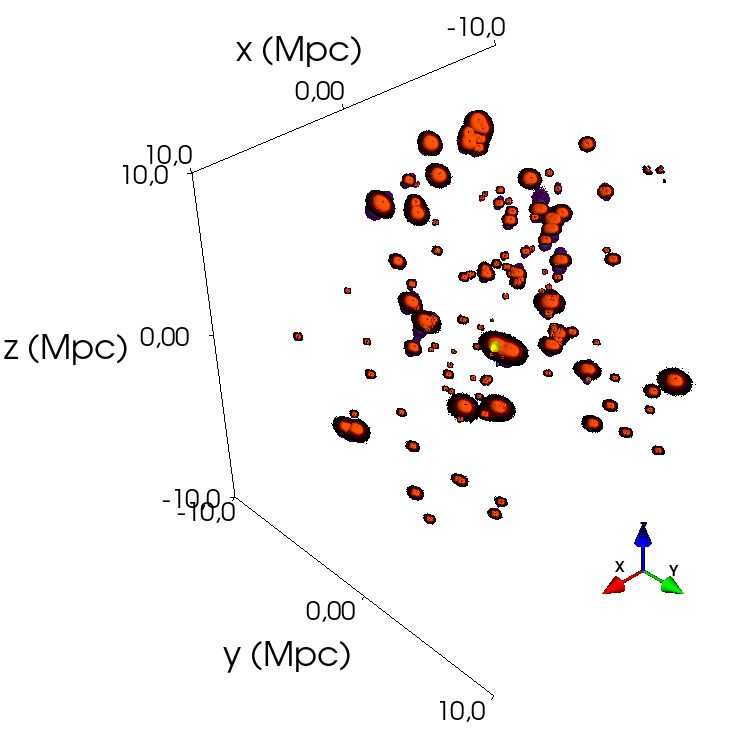}
   \caption{Capture of a 3D contour plot of the PDM density of the Local Volume when distant clusters from Table~\ref{table:external_sources} are taken into account. The MW is the yellow ball at the center. The xy plane at $z=0$ is the MW disk plane. Color bars are the same as in Figure~\ref{fig:10Mpc_everything}. Because of projection effects, and because halos have less PDM at a given radius compared to the previous case due to the stronger EFE, we see almost exclusively negative PDM zones on this capture (oriented towards Virgo, located at $(x,y,z)=(0.38,2.26,17.60)$ Mpc).}
\label{fig:3D_everything}
\end{figure*}

With those new contributions, the PDM landscape  changes drastically as can be seen on the plane cuts on Figure~\ref{fig:10Mpc_everything}, or the 3D visualization in Figure~\ref{fig:3D_everything}.  In this configuration, the Virgo supercluster, of which we place the centre at the position (375,2260,17600)~kpc, is the dominant EFE source and thus all negative zones of PDM are perpendicular to its direction. Furthermore, this influence brings much less diversity as a lot of galaxy-galaxy interactions between sources in the Local Volume observed on Figure~\ref{fig:10Mpc_no_ext_sources} and Figure~\ref{fig:3D_no_ext_sources} are smoothed away. Due to a stronger EFE, PDM halos do not extend as far as in the previous case and are less massive, a point we inspect further in the next subsection. 

Let us note here that modeling things this way leads us to having galaxy clusters as the only sources of gravity outside the Local Volume. This could certainly be an issue at the scale of the most distant clusters, in terms of average density in particular, because underdensities are {\it de facto} neglected: these could lower the EFE strength and change its direction. However, with Virgo being by far the dominating source of EFE for the Local Volume, and with it being so close (17.8 Mpc), our assumptions should cause no issue at the scale of the Local Volume.

Finally, as an interesting alternative possibility (if we imagine for instance that the residual missing mass in MONDian clusters does not contribute to the EFE), we show the same computation when only the baryonic mass from clusters is taken into account in the Appendix (Figure~\ref{fig:ext_sources_baryonic_mass}). The EFE from distant sources still dominates in this case, albeit less overwhelmingly.

\subsection{SHMR}
\label{subsec:SHMR}
When a galaxy is embedded in an external field, its total PDM mass is finite. If the external field is constant over the size of the galaxy, the total mass is given by equation (57) of \citet{2010MNRAS.403..886M}. However, this does not compare well to the virial mass of DM halos, usually defined at $r_{200}$. \citet{2015MNRAS.446..330W} have computed a SHMR for MOND, both isolated and with EFE, using analytical formulas, and predict truncation radii in PDM halos due to the EFE, leading to less enclosed mass. We will also derive a SHMR based on our sample, but our computation herafter will be purely numerical.

In order to compute the individual PDM mass within $r_{200}$ for galaxies in our sample, we make a zoomed computation of the PDM density around each galaxy with resolution 1 kpc by growing a sphere around it until a mean density of $200\rho_{\rm crit}$ is reached. The result can be found in the Appendix (Table~\ref{table:local_sources}). Some galaxies too embedded into the halo of another neighbouring galaxy for it to make sense to compute their PDM mass were excluded from the computation. This includes mostly small satellite galaxies of massive hosts, but also some close systems such as M81 and M82. For this computation, the EFE from the distant sources of Table~\ref{table:external_sources} is correctly taken into account.

\begin{figure*}
\centering
  \includegraphics[angle=0,  clip, width=15cm]{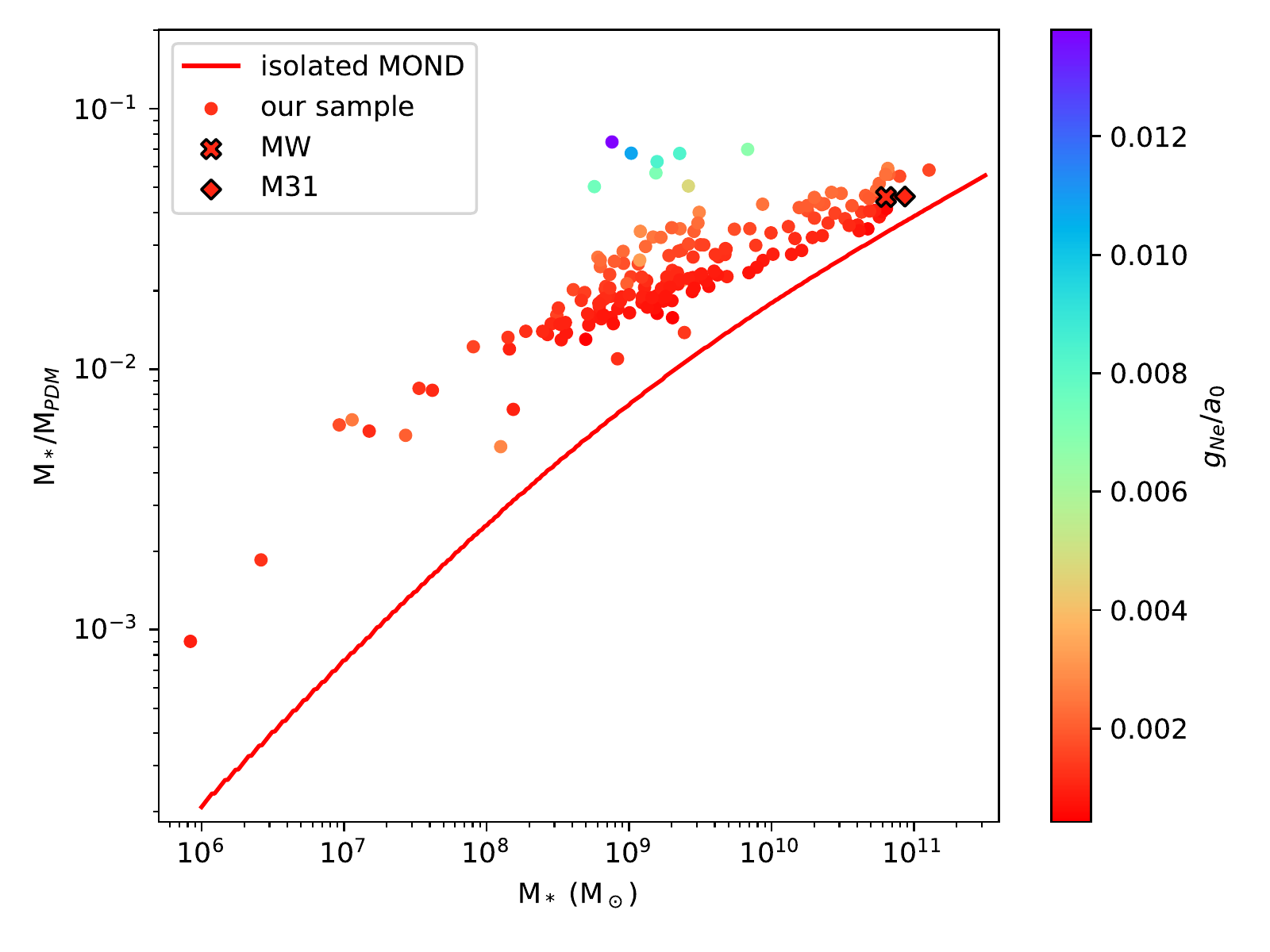}
   \caption{Stellar mass over PDM mass (at a computed $r_{200}$) as a function of stellar mass. Each coloured dot is a galaxy from Table~\ref{table:local_sources}. The color bar indicates the Newtonian external gravitational acceleration $g_{Ne}$ at the location of each galaxy in units of $a_0$. The MW is represented by a cross and M31 by a diamond. The red curve is the analytically computed PDM mass at $r_{200}$ for the isolated MOND case. We notice as expected because of the EFE a vertical gradient in terms of $g_{Ne}$, explaining the scatter at a given stellar mass.}
\label{fig:stellar_mass_relation_external}
\end{figure*}

\begin{figure*}
\centering
  \includegraphics[angle=0,  clip, width=15cm]{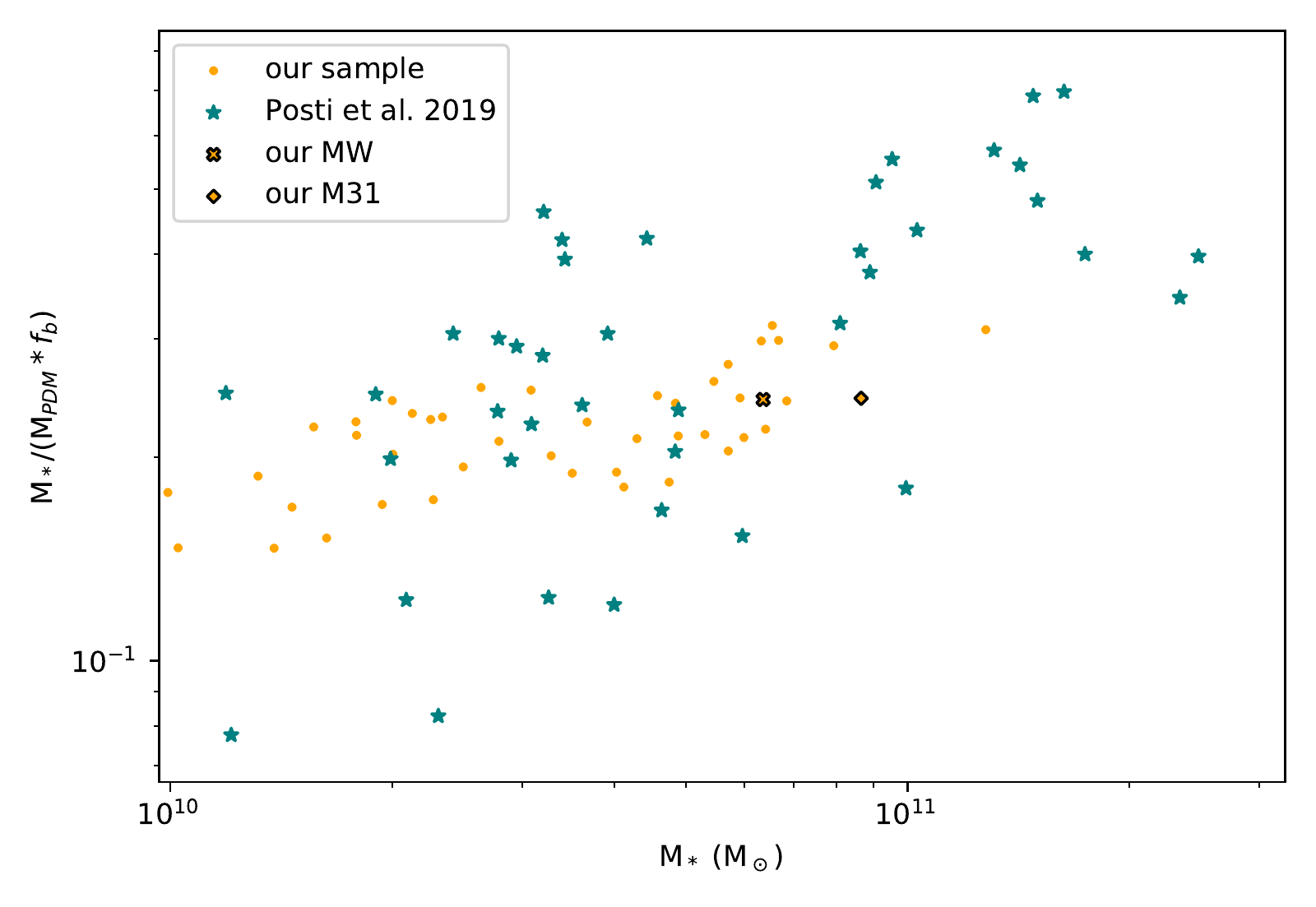}
   \caption{Stellar mass over PDM mass (at a computed $r_{200}$) divided by the cosmological baryon fraction $f_b=0.188$ as a function of stellar mass. Zoom on the most massive galaxies. Each orange dot is a galaxy from Table~\ref{table:local_sources}. The MW and M31 from our computation are respectively the cross and the diamond. Star symbols represent fits of \citet{2019A&A...626A..56P}, in which a halo mass is estimated in the $\Lambda$CDM context for SPARC galaxies \citep{2016AJ....152..157L}.}
\label{fig:shmr_vs_posti_2019}
\end{figure*}

Figure~\ref{fig:stellar_mass_relation_external} now shows the MOND ``stellar-to-PDM" relation where each galaxy of our catalogue is represented by a dot coloured according to the Newtonian external gravitational acceleration $g_{Ne}$ at their location, and a lower limit can be found as the red curve representing the isolated MOND situation as in Subsection~\ref{subsec:isolated_MOND}. The broad range of external gravitational accelerations for different galaxies has two effects compared to the isolated MOND case: (i) it induces a PDM mass loss in galaxies, with the effect being more pronounced in the less massive ones, and (ii) it creates some scatter for a given stellar mass, as can be seen by the vertical gradient in colour representing $g_{Ne}$. Globally, the shape of the SHMR is well represented by a single power-law (hence a straight line in log-log) with scatter 0.14 dex. Note that, since we neglected in this case the scatter on the stellar-to-gas mass relation, the true scatter should be larger. The single power-law behaviour is similar to the behaviour of the SHMR that has been found for disk galaxies in the $\Lambda$CDM context, in disagreement with abundance matching expectations \citep[e.g.][]{2019A&A...626A..56P, 2019A&A...629A..59P}, and has been attributed to a morphologically-dependent SHMR by \citet{2021A&A...649A.119P}. However, in the MOND context, morphology should (in principle) play no role in the shape of the SHMR, so the only explanation in this case would be an environmental one, either through varying degrees of EFE, or with a higher mass discrepancy for early-type galaxies residing in groups and clusters. Concerning massive late-type galaxies, we show as an illustration in Figure~\ref{fig:shmr_vs_posti_2019} a zoom on our MONDian SHMR for galaxies with stellar mass greater than $10^{10}$ M$_\odot$, where a tension with abundance matching was found in the standard context, and we compare them with data from Figure 2 of \citet{2019A&A...626A..56P}. The global agreement is striking.

We also find that both the MW and M31 have a dynamical mass below what is expected from a $\Lambda$CDM stellar-to-halo mass relation, in agreement with \citet{2021RNAAS...5...23M}. Indeed, we find PDM halo masses at $r_{200}$ of $1.39\times10^{12}$ M$_\odot$ for the MW and $1.88\times10^{12}$ M$_\odot$ for M31, while abundance matching instead predicts a halo mass of approximately $2.5\times10^{12}$ M$_\odot$ for our stellar mass of the MW ($6.37\times10^{10}$ M$_\odot$) and approximately $6\times10^{12}$ M$_\odot$ for our stellar mass of M31 ($8.65\times10^{10}$ M$_\odot$), as can be seen on Figure 1 of \citet{2021RNAAS...5...23M}.

\begin{figure*}
\centering
  \includegraphics[angle=0,  clip, width=15cm]{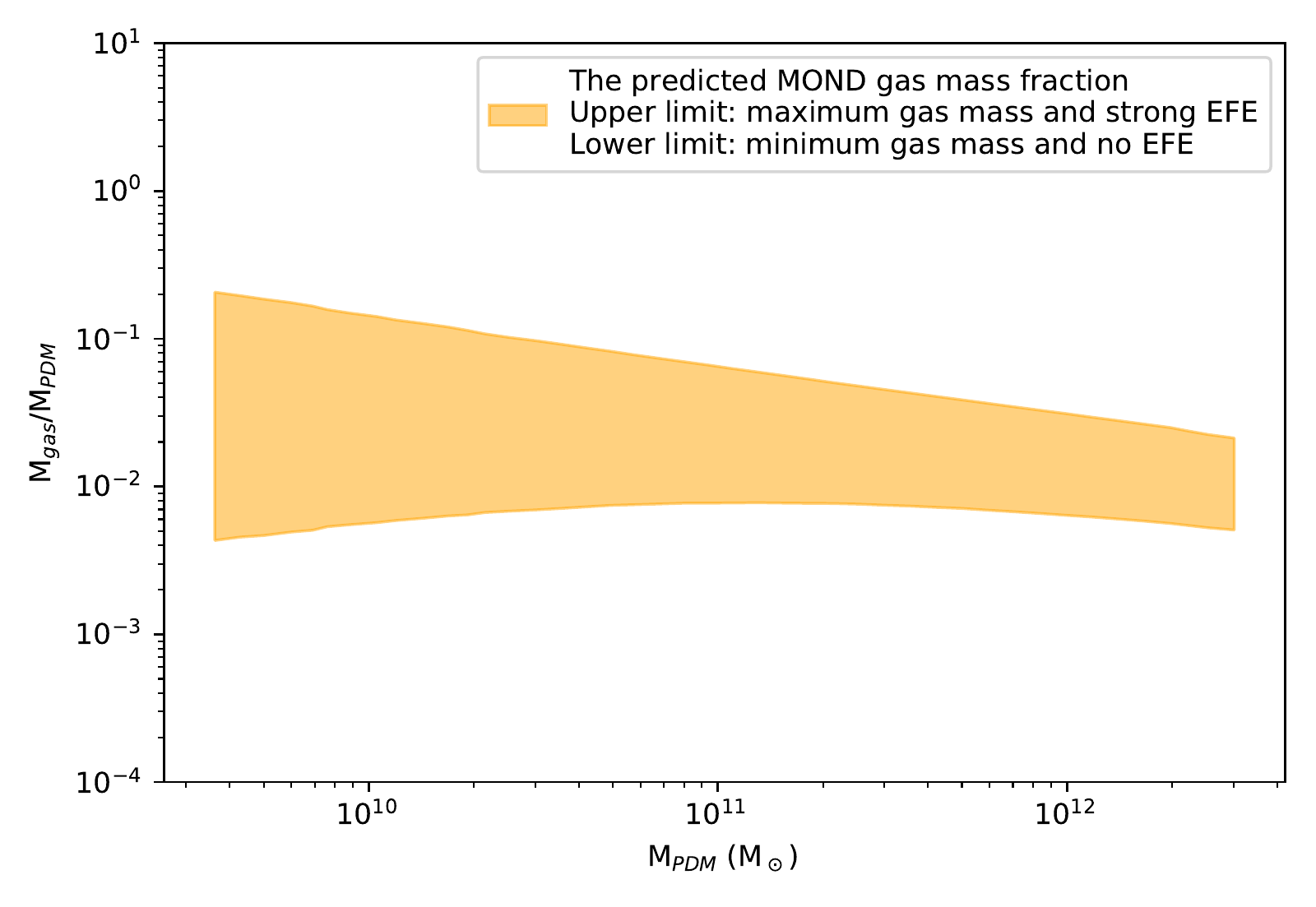}
   \caption{Gas mass over PDM mass (at a computed $r_{200}$) as a function of PDM mass. The shaded area represents the expected values, with the isolated MOND case combined with the lowest gas mass estimation from Eq~\eqref{eq:stellar_mass_to_H} as a lower limit, and a strong EFE induced by a $g_{Ne}=0.02$ $a_0$ combined with the highest gas mass estimation from Eq~\eqref{eq:stellar_mass_to_H} as the upper limit. The gas-to-halo mass relation predicted by MOND remains flat.}
\label{fig:GHMR}
\end{figure*}

Finally, we explore again the question of the gas-to-halo mass relation in the case of an EFE, which was found to be flat in the isolated MOND case. On Figure~\ref{fig:GHMR}, we update this figure by using as a lower limit the isolated MOND computation of Subsection~\ref{subsec:isolated_MOND} combined with the lowest possible gas mass from Eq~\eqref{eq:stellar_mass_to_H}, and as an upper limit, the PDM mass at $r_{200}$ of test galaxies of a range of stellar masses and the highest possible corresponding gas mass from Eq~\eqref{eq:stellar_mass_to_H}, under a $g_{Ne}=0.02$ $a_0$ EFE, which is the typical maximum EFE that can be reached in the volume occupied by the SPARC \citep{2016AJ....152..157L} galaxies, larger than the maximum EFE in the Local Volume. As can be seen on this figure, the inclusion of the EFE has strongly increased the scatter of the gas-to-halo mass relation predicted by MOND, but the overall shape remains very flat.

\subsection{The average PDM density in the Local Volume}

\citet{2018AN....339..615K} find, in the context of $\Lambda$CDM, a total mass of $10^{14}$ M$_{\odot}$ for the volume enclosed in a sphere of radius 11 Mpc centred on the MW. For this, they simply add the halo masses expected from the SHMR around all galaxies of the UNGC catalog. On Figure~\ref{fig:omegas}, we add on the plot the mean PDM density as a function of distance to the MW for our sample of galaxies (orange curve). We then extrapolate this to the whole UNGC (black curve) by using a linear fit to the SHMR of Figure~\ref{fig:stellar_mass_relation_external} to obtain the PDM mass at $r_{200}$ as a function of stellar mass of the rest of the galaxies in the catalogue. Note that this indirectly includes a gas component since the PDM mass of the galaxies in our sample was computed with an added gas mass to the stellar mass. The difference in $\Omega_{\rm pdm}$ between our sample and the extrapolation to the whole UNGC is slim, showing that the galaxies we did not select do not contribute much mass to the Local Volume. For our 206 galaxies in MOND, we find a total mass of $5.74\times10^{13}$ M$_\odot$ for PDM and of $6\times10^{13}$ M$_\odot$ when including baryons, corresponding to a mean PDM density (in galaxies only) $\Omega_{\rm pdm} = 0.07$. By extrapolating our SHMR of the previous subsection, knowing that it overestimates the PDM masses of galaxies which would reside in high external field environments for which no PDM mass was computed, we find an upper limit to the PDM mass enclosed in galaxies of the Local Volume, $\Omega_{\rm pdm} = 0.078$, hence much lower than the average CDM density in a $\Lambda$CDM Universe but close to the $\Omega_{DM} \approx 0.1$ corresponding to the mass of $10^{14}$ M$_{\odot}$ deduced by \citet{2018AN....339..615K}.

\section{A fiducial MOND model for the Milky Way in its environment}
\label{sec:MW}

This model of the Local Volume offers us a unique opportunity to produce a fiducial MOND model of the Milky Way when actually embedded in its gravitational environment.

Solving the QUMOND Poisson equation in~\eqref{eq:MOND_poisson} with appropriate boundary conditions gives us the MOND potential. We do it here for a cube of 1 Mpc side length centered on the MW, comprising 500 identical cells inc each dimension, for a resolution of 2 kpc. Our Poisson solver is based on a Gauss-Seidel iterative process with acceleration by over relaxation (which introduces a relaxation parameter, making the Gauss-Seidel iterative process converge faster). It operates on a grid with cells of identical size (no adaptive refinement) and gives a gravitational potential value for each vertex. 

\subsection{Computing the PDM density}
\label{subsec:computing_MW_PDM}

The first step towards solving the QUMOND Poisson equation is to solve the baryonic Poisson equation $\Delta \Phi_b = 4\pi G\rho_b$ to obtain the baryonic potential $\Phi_b$ of the MW, since we do not use an analytical potential. For this computation, the MW is modeled as an exponential disk galaxy following Section 2.7 of \citet{2008gady.book.....B}, i.e. with density profiles for a bulge component, for a thin and a thick disk, and for a gas corona. We switch from the less realistic Miyamoto-Nagai profile of Subsection~\ref{subsec:UNGC} since an analytical formula is not required anymore. The disks have a scale length of 2 kpc, and scale heights of 0.3 kpc and 1 kpc for the thin and thick disks respectively. The gas corona has a scale radius of 4 kpc, and a hole of 4 kpc radius at its center. The density parameters have been chosen to numerically give a total baryonic mass of $7.5\times10^{10}$ M$_\odot$ for the MW, of which $6.37\times10^{10}$ M$_\odot$ is in the disk (including $1.13\times10^{10}$ M$_\odot$ in gas), and $7\times10^{9}$ M$_\odot$ is in the bulge. The boundary condition used for the baryonic Poisson equation is $-GM_b/r$ where $G$ is the gravitational constant, $M_b$ is the baryonic mass of the MW, and $r$ is the distance from the MW. 

\begin{figure*}[!htb]
\centering
  \includegraphics[angle=0,  clip, scale=0.38]{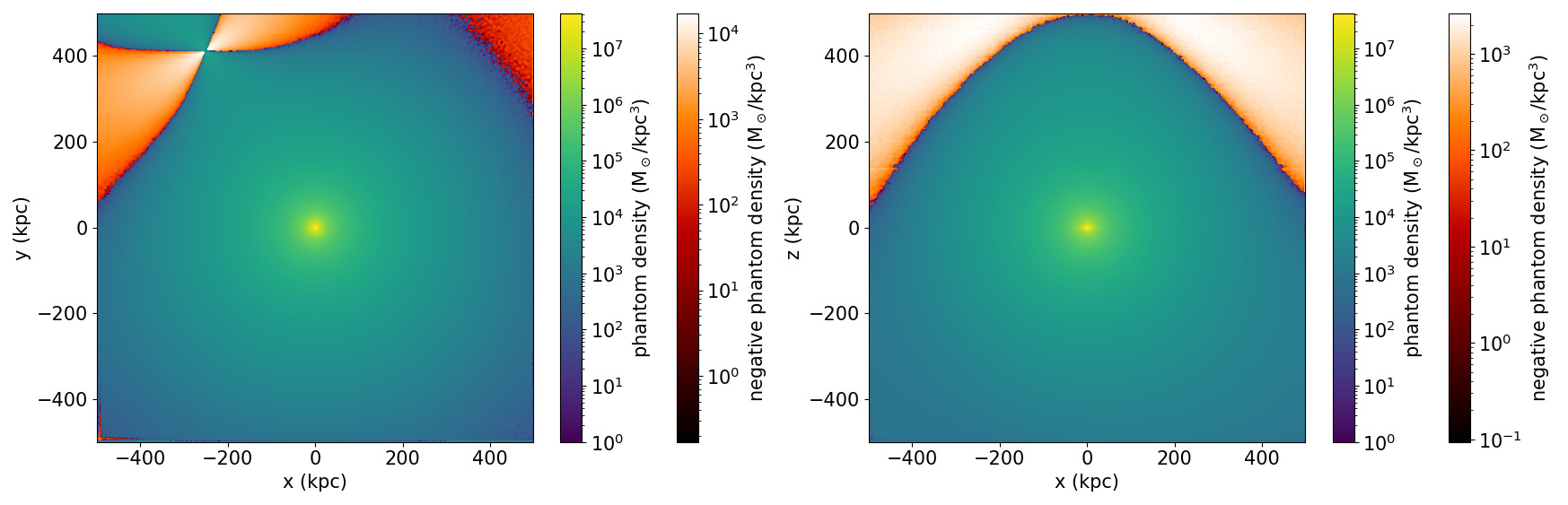}
   \caption{Phantom dark matter density map around the MW. Plane cuts. \textit{Left panel:} galactic plane z=0. Zones of negative PDM density due to M31 at the position $(x,y,z)=(-383,619,-286)$ kpc (top-left corner) and Virgo at the position $(x,y,z)=(0.38,2.26,17.60)$ Mpc (top-right corner). \textit{Right panel:} edge-on view y=0. Zone of negative PDM density due to Virgo.}
\label{fig:phantom_MW_500kpc_everything}
\end{figure*}

The second step towards solving the QUMOND Poisson equation is to compute the PDM density $\rho_{\rm PDM}$, which is done using Equation~\eqref{eq:rho_ph_vertex} as in Subsection~\ref{subsec:QUMOND}. The Newtonian potential $\Phi_N$ used for this computation includes the baryonic potential of the MW obtained from the previous Poisson integration above, as well as the baryonic potential of all the sources used for the PDM density computation in Section~\ref{sec:mapping}, i.e. M31, plus all the Local Volume galaxies of Table~\ref{table:local_sources}, and all the external sources from Table~\ref{table:external_sources}. 
The result can be seen on Figure~\ref{fig:phantom_MW_500kpc_everything}. The most notable features are the negative PDM areas. The large composite negative PDM area on the xy plane is the consequence of the EFE from M31 (at the position $(x,y,z)=(-383,619,-286)$ kpc) and the EFE from the Virgo supercluster (at the position $(x,y,z)=(0.38,2.26,17.60)$ Mpc); the one on the xz plane above the MW is the consequence of the EFE from the Virgo supercluster. We note that those negative PDM areas, which arise when the internal and external gravitational accelerations are comparable, are far from the center of the MW. It would be extremely interesting to be able to probe precisely those very outer regions of the stellar halo near the virial radius of the PDM halo, or to analyze the behaviour of what remains of the hot gas corona in these outer regions.

Before computing the MOND potential, we run a PDM density computation of the same volume without including the baryonic potentials of the MW and the LMC in order to obtain the background information. The PDM density from this background is subtracted from the full PDM density. This allows us to isolate the PDM from the MW and from the LMC while still taking into account the background field they are in. Note that this is not a perfect solution: negative PDM areas will be exacerbated by this process because we will take away from them some positive background PDM (since without the MW and the LMC, there is no negative PDM in the same area). We found however that it is a reasonable solution to a problem that is hard to solve because of its non-linearity. 

This special treatment of the LMC is justified by its importance in the MW system. It would not make much sense to try to estimate its current PDM mass as it is now embedded in the halo of the MW and under a very strong EFE caused by its host. However, from its stellar mass of $1.78\times10^{9}$ M$_\odot$ (derived from the UNGC and our process of Subsection~\ref{subsec:UNGC}), we can estimate its PDM mass prior to its infall into the MW. In the isolated MOND case, we can use Eq~\eqref{eq:isolated_PDM_mass}, giving a PDM mass of $1.95\times10^{11}$ M$_\odot$. More realistically, using an extrapolation of our SHMR of Subsection~\ref{subsec:SHMR} and thus taking into account the EFE on the Local Universe, we find a PDM mass of $8.14\times10^{10}$ M$_\odot$ prior to infall. Such a high PDM mass should have a noticeable effect in the response of the stellar halo of the MW to the LMC infall \citep[see e.g.]{2019ApJ...884...51G}, in addition to creating interesting negative PDM density zones at infall. Studying the detailed response of the stellar halo of the MW to the LMC infall in MOND will be the topic of further work, using the fiducial model of the MW presented here as a backbone.

\subsection{Computing the MOND potential of the Milky Way}
\label{subsec:MW_MOND_potential}

Equation~\eqref{eq:MOND_poisson} is solved with source terms $\rho_b$ being the sum of all the baryonic density profiles for the MW described in Subsection~\ref{subsec:computing_MW_PDM}, $\rho_{\rm PDM}$ being the PDM density for the MW obtained after the background substraction described in Subsection~\ref{subsec:computing_MW_PDM}, and with boundary condition $-GM/r$ where $M$ is the sum of the baryonic and PDM masses for the MW obtained by numerical integration of the density in our 1 Mpc side-length box. As expected from the PDM density plot of Figure~\ref{fig:phantom_MW_500kpc_everything}, no strong asymmetry is noticeable in the derived QUMOND potential, a sign that the MW is not under a strong EFE.

\begin{figure*}
\centering
  \includegraphics[angle=0,  clip, width=15cm]{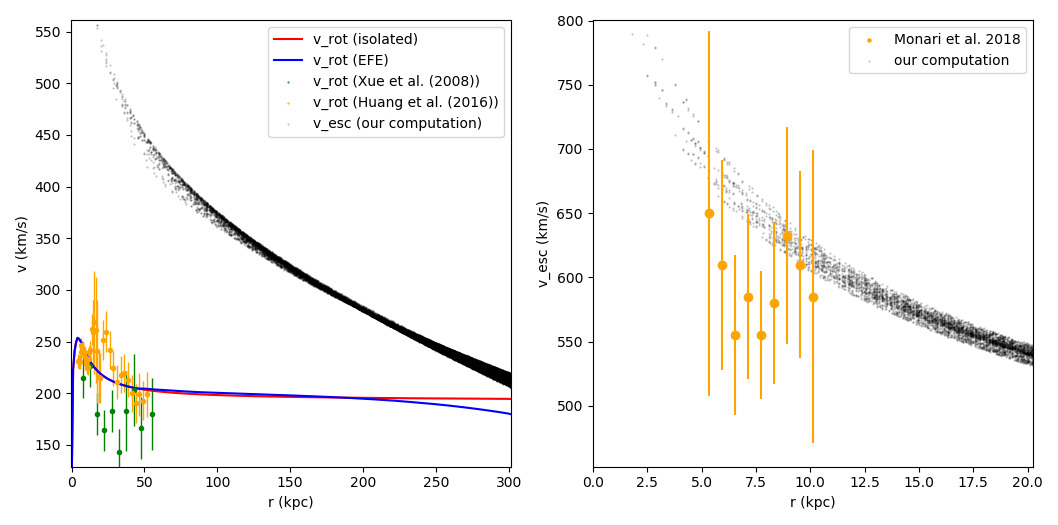}
   \caption{\textit{Left panel:} Escape speed and circular velocity as a function of radius for the MW in QUMOND. Black dots are our escape speed computation. The vertical scatter, more pronounced after 200 kpc, is a consequence of the EFE. The red curve is the rotation curve in the case of isolated MOND, and the blue curve is the rotation curve with EFE. Green points are from \citet{2008ApJ...684.1143X} and orange points are from \citet{2016MNRAS.463.2623H}. \textit{Right panel:} Zoom on the inner 20 kpc for the escape speed computation. The bigger orange dots are data points from \citet{2018A&A...616L...9M}.}
\label{fig:escape_speed}
\end{figure*}

The escape speed $v_{esc}$ for the MW is computed from the MOND potential $\Phi$ via the following formula:
\begin{equation}
\label{eq:escape_speed}
v_{esc} = \sqrt{2(\Phi_\infty-\Phi)}
\end{equation}
where $\Phi_\infty$ is the weakest potential in our volume. The result of this computation can be seen in Figure~\ref{fig:escape_speed}. A zoom on the inner 20 kpc can be seen on the right panel, showing a reasonable agreement with the data from \citet{2018A&A...616L...9M}. This is in line with our PDM halo mass of $1.39\times10^{12}$ M$_\odot$, similar to the estimate of the total mass by \citet{2018A&A...616L...9M}. Furthermore, we compute the rotation curves for the MW in the isolated MOND and EFE cases. The acceleration $g$ is extracted by taking the norm of the gradient of the MOND gravitational potential. Then, the mean acceleration $g(r)$ at each radius $r$ is computed. The circular velocity $v_{rot}$ at a radius $r$ is obtained via 
\begin{equation}
\label{eq:v_rot}
v_{rot}(r) = \sqrt{g(r)r}.
\end{equation}
The result can be seen on the left panel of Figure~\ref{fig:escape_speed}. The change in curvature induced by the EFE on the blue curve happens between 150 and 200 kpc and separates it from its asymptotically flat isolated MOND counterpart (red curve). As a comparison to observations, we add data points from \citet{2008ApJ...684.1143X} and \citet{2016MNRAS.463.2623H}. In Subsection~\ref{subsec:NGC5055}, we take a look at this effect on the rotation curve in a more drastic case, with a less massive galaxy under a stronger EFE. 

\section{Dynamical and lensing signatures in galaxies under a strong EFE}
\label{sec:signatures_of_EFE}

\subsection{NGC~5055: an archetypical galaxy \\ under a strong EFE}
\label{subsec:NGC5055}

In a recent study, \citet{2020ApJ...904...51C} have found that galaxy rotation curves in the SPARC database \citep{2016AJ....152..157L} were significantly better fitted in MOND when including a contribution from the EFE, which affects the few outermost observed datapoints. While a few extreme cases indicated that a stronger EFE was needed in gravitational environments that are effectively stronger, the global correlation was still weak. This might be because it is difficult to translate the large-scale gravitational accelerations expected in a $\Lambda$CDM Universe into those expected in a MOND Universe. Therefore, using the QUMOND formulation would certainly allow to probe more clearly the possible correlation between the EFE needed to fit galaxy rotation curves and the true EFE in which galaxies reside. 

In our sample, NGC5055 is a galaxy in the strong gravitational influence of the Virgo supercluster with $g_{Ne}=2.7\times 10^{-3}\ a_0$ at its location, and it is also present in the SPARC database. It therefore represents an interesting archetypical galaxy residing in a strong external field.

We compute and compare the rotation curve of NGC5055 both for the MOND isolated case, and for the case where all the Local Volume sources from Table~\ref{table:local_sources} and external sources from Table~\ref{table:external_sources} are included and cause an EFE. In our Local Volume cube, NGC5055 lies 8.99 Mpc away from the MW at the position $(-670,2340,8654)$~kpc. Among our selected galaxies, it is one of the closest to the Virgo supercluster, and thus one of the most affected by the EFE coming from it.

The galaxy is modeled using an exponential disk profile with an effective radius of 4.18 kpc, a disk scale length of 3.2 kpc, and a baryonic mass of $5.48\times 10^{10}$ M$_\odot$. Note that this is a mass close to that of the best fitting model of \citet{2020ApJ...904...51C} which we use here for the sake of comparison, although their mass is a bit lower than the mass we obtain via the process detailed in Subsection~\ref{subsec:UNGC}. This way, the model chosen here is in principle guaranteed to give a good representation of the data.

\begin{figure}
\centering
  \includegraphics[angle=0,  clip, width=8cm]{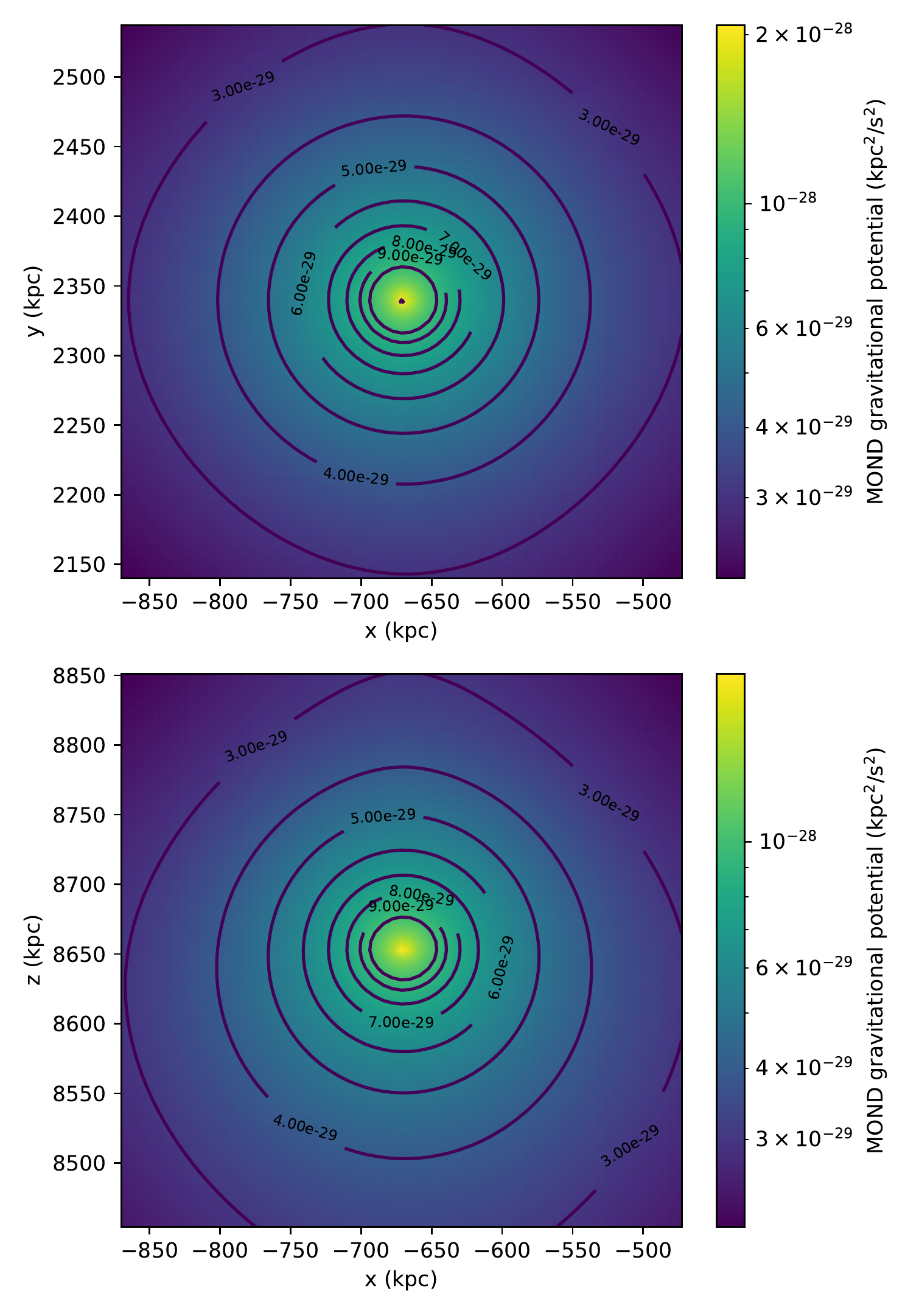}
   \caption{QUMOND gravitational potential of NGC5055 in the case of an EFE, with the egg-shaped contours visible. Plane cuts. \textit{Top panel:} galactic plane z=0. \textit{Bottom panel:} edge-on view y=0. The values on the axes are still centered on the MW position.}
\label{fig:NGC5055_potential}
\end{figure}

In order to recover the MOND potential in the case where all sources are included, a first integration of Equation~\eqref{eq:MOND_poisson} is done in a cube of 1 Mpc side length centred on NGC5055 with a resolution of 2 kpc. The boundary condition is $-GM/r$ where $G$ is the gravitational constant, $M$ is the sum of the baryonic and PDM masses for NGC5055, and $r$ is the distance to NGC5055. The computed PDM mass in this volume is $1.43\times 10^{12}$ M$_\odot$. Then a refined integration of Equation~\eqref{eq:MOND_poisson} is done in a smaller cube of 400 kpc side length with a resolution of 800 pc, with boundary condition extracted from the bigger cube. The computed PDM mass in this volume is $9.08\times 10^{11}$ M$_\odot$. The resulting potential can be seen in Figure~\ref{fig:NGC5055_potential}. The  asymmetry caused by the strong EFE can be seen on the xz plane cut. This potential in the shape of an egg oriented in the direction of the dominating EFE is typical, and already showcased in e.g. \citet{2017A&A...603A..65T,2018A&A...609A..44T} for the case of a satellite under the EFE of its host. Similarly, lopsidedness of the potential has been investigated by \citet{2017ApJ...844..130W} in the case of galaxies in clusters. The effect is however mild, and it is therefore not clear that a direct dynamical detection of this asymmetric potential would ever be possible for such a disk galaxy residing in a strong EFE. Dynamically, the best we can probably hope to achieve is the Keplerian decline of the rotation curve associated to the EFE. With this in mind, we will now study in detail a new formula proposed in \citet{2021arXiv210904487F} and compare it to our exact QUMOND calculations.

\begin{figure}
\centering
  \includegraphics[angle=0,  clip, width=8cm]{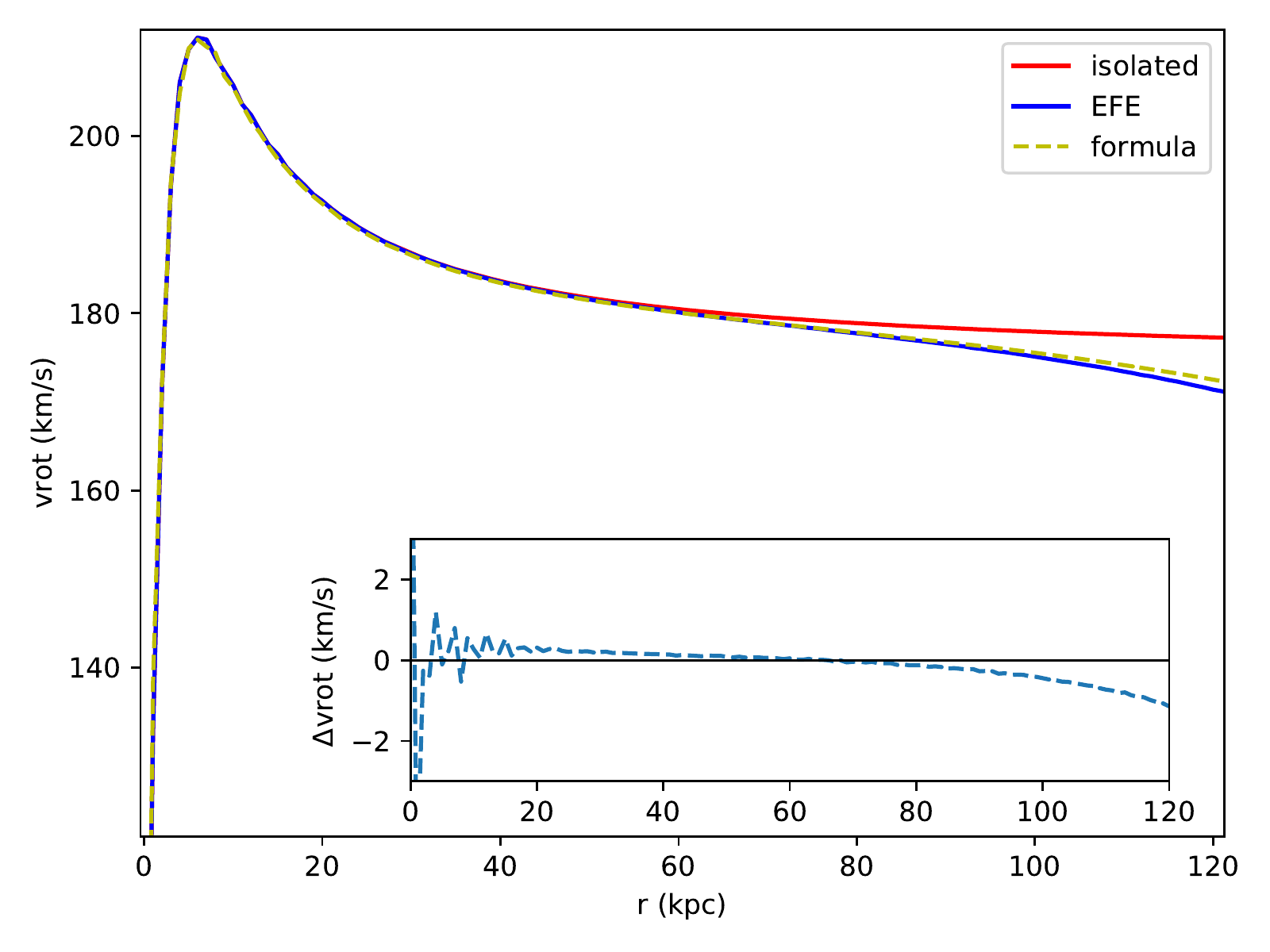}
   \caption{Computed rotation curves for NGC5055 with or without external field effect. The isolated MOND case is in red; the EFE case with all the sources from Table~\ref{table:local_sources} and Table~\ref{table:external_sources} is in blue; the yellow dashed curve is the one given by the analytical formula of Eq~\eqref{eq:gr_formula}. \textit{Inset panel:} difference between the blue curve of the EFE case and the yellow dashed curve of the analytical formula of Eq~\eqref{eq:gr_formula}.}
\label{fig:NGC5055_rot_curve}
\end{figure}

The rotation curves, computed as described in Subsection~\ref{subsec:MW_MOND_potential}, are presented in Figure~\ref{fig:NGC5055_rot_curve}. The red and blue curves, respectively of the isolated and EFE cases, are extremely close up to a radius of approximately 55 kpc where a change of slope is noticeable, with the red curve of the isolated case staying more or less flat while the EFE kicks in and brings the blue curve down. 

 We note that the rotation curve with EFE is very well fit by the formula proposed by \citet{2021arXiv210904487F}, approximating the average radial acceleration $g_r(r)$ over a sphere at a given radius $r$ in a constant external field :
\begin{equation}
\label{eq:gr_formula}
g_r(r) =
\begin{cases}
      \nu\left(\frac{g_{Ni}(r)+\frac{g_{Ne}^2}{3g_{Ni}(r)}}{a_0}\right)g_{Ni}(r),& \text{if}\ g_{Ni}(r) \geq g_{Ne} \\
      \nu\left(\frac{g_{Ne}+\frac{g_{Ni}(r)^2}{3g_{Ne}}}{a_0}\right)g_{Ni(r)},& \text{if}\ g_{Ne} \geq g_{Ni}(r) 
\end{cases}
\end{equation}
where $\nu$ is the transition function of Eq~\eqref{eq:nu_function}, $g_{Ni}(r)$ is the Newtonian internal acceleration at radius $r$ from the baryonic profile of the system studied, and $g_{Ne}$ is the (constant) Newtonian external acceleration. 
The rotation curve obtained from this formula when applied to NGC5055 with our parameters is the yellow dashed curve on Figure~\ref{fig:NGC5055_rot_curve}, with the difference relative to the EFE curve being shown in the inset panel. The description provided by this formula is impressive, and only starts noticeably departing from the numerically computed EFE curve at a radius of approximately 80 kpc, where it starts to underestimate the EFE. We thus advocate to use this formula in future studies of the EFE in a QUMOND context.

\subsection{The concave-lens signature~\\~of the negative PDM zones}

We have seen that the gravitational potential of a galaxy like NGC5055 under a strong EFE does display an egg-shaped asymmetry. It is however not clear that it would be easily detectable dynamically, and does not in itself represent a smoking gun of negative PDM zones. A perhaps better indicator could however be obtained via gravitational lensing from a large amount of such galaxies located in strong gravitational field environments. 

\begin{figure}
\centering
  \includegraphics[angle=0,  clip, width=8cm]{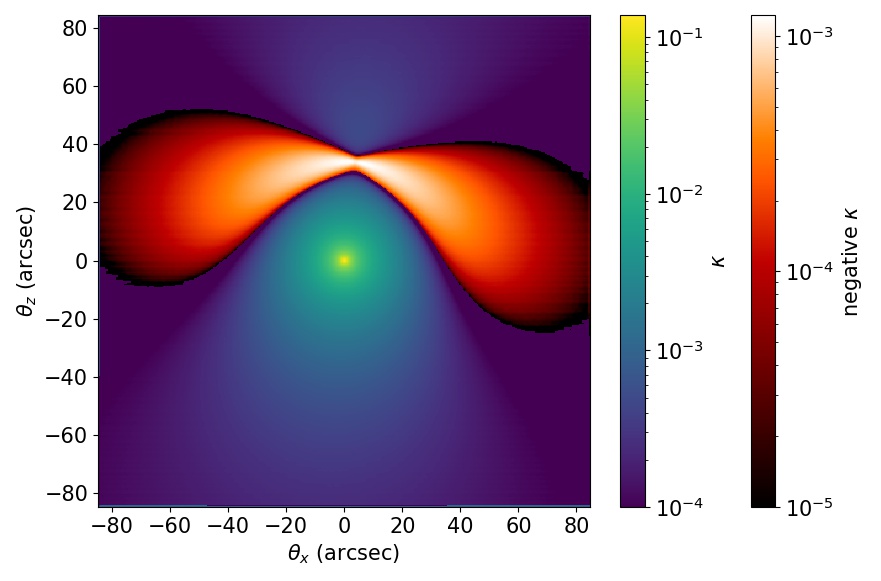}
   \caption{Convergence parameter $\kappa$ for an NGC5055-like galaxy centred at $\theta_x=\theta_z=0$ under the EFE of Virgo as in our study, seen edge-on, used as a gravitational lens placed at $z = 0.3$ for sources at $z = 5$.}
\label{fig:lensing}
\end{figure}

In order to investigate this, we computed the surface density of NGC5055 seen from various angles, which we show can be negative due to the EFE induced by the Virgo supercluster. This is possible if we look at the galaxy edge-on (i.e., observing it perpendicular the the EFE direction) but not face on (i.e., observing it along the EFE direction), in which case the PDM halo largely outweighs the negative PDM density. We then artificially place this system at a redshift $z = 0.3$ and see if it can act as a concave diverging lens for sources at $z = 5$. We compute the convergence parameter 
\begin{equation}
\label{eq:kappa}
\kappa = \frac{1}{2}\nabla^2\Upsilon
\end{equation}
where $\Upsilon$ is the deflection potential, directly linked to the surface density and the distances of the lens and the source, as in \citet[equation 110]{2012LRR....15...10F}. The result of this computation can be seen in Figure~\ref{fig:lensing}. The convergence parameter $\kappa$ reaches negative values of the order of $-10^{-3}$ in this case. This plot also highlights the double-bottleneck shape of PDM around the negative area which could already be noticed on the 3D plots (Figure~\ref{fig:3D_no_ext_sources}, Figure~\ref{fig:3D_everything}). 

Keeping the same distances for lens and source, we also considered a more extreme theoretical case of a galaxy of baryonic mass $5\times10^{10}$ M$_\odot$ under an EFE of $g_{Ne}=0.02$ $a_0$, a high value that we already used in Subsection~\ref{subsec:SHMR} as an upper limit. We find in this configuration that the convergence parameter $\kappa$ can reach negative values of $-3\times10^{-3}$. The phenomenon of lensing with negative convergence parameter has already been studied in e.g. \citet{2013PhRvD..88b4049I} and \citet{2014PhRvD..90h4026N}, where it was argued that it should produce radially distorted images.

\begin{figure*}
\centering
  \includegraphics[angle=0,  clip, width=15cm]{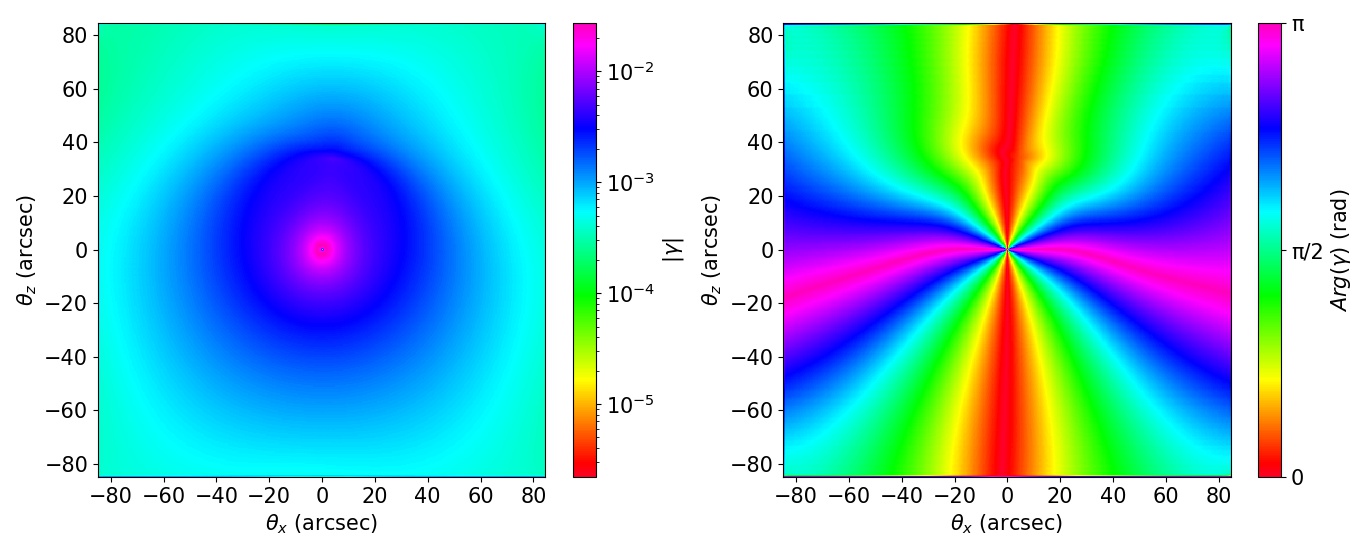}
   \caption{Shear vector $\gamma$ corresponding to the convergence map of Fig.~\ref{fig:lensing}. \textit{Left panel:} shear amplitude $|\gamma|$. \textit{Right panel:} shear angle ${\rm Arg}(\gamma)$.}
\label{fig:shear_EFE}
\end{figure*}

 Here, from the convergence parameter, we compute the shear $\gamma=\gamma_1+i\gamma_2$ as in \citet[Eq. 13]{2013arXiv1312.5981H}, and plot the resulting shear amplitude $|\gamma|=\sqrt{\gamma_1^2+\gamma_2^2}$ and shear angle ${\rm Arg}(\gamma)$ in Figure~\ref{fig:shear_EFE}. We observe a strong asymmetry on the shear amplitude following the NGC5055-Virgo axis, with a feature around ($\theta_x=0,\theta_z=40)$ arcsec, the position (when placing the galaxy at $z=0.3$) of the peak intensity of the negative PDM caused by the EFE. Perhaps more strikingly, the shear angle map is heavily distorted at this same position around ($\theta_x=0,\theta_z=40)$ arcsec. As a means of comparison, we plot in Figure~\ref{fig:shear_isolated} (Appendix) the shear map for the same configuration but in the isolated MOND case (i.e. no EFE).
 
 Such a negative PDM zone signature cannot in principle be distinguished from the effect of a prominent underdensity in the standard context, but a statistical analysis correlating them with the direction of the expected EFE would be a smoking gun for MOND. This is something that could potentially be detected only statistically in weak galaxy-galaxy lensing. Stacking lenses that are in strong EFE environments near large scale structures or galaxy clusters could perhaps allow such a detection: the correlation of the location of the drop in $\kappa$ with the large-scale environment would be the key. We leave it to further work to develop mock catalogues of negative convergence maps expected in MOND, their associated shear, and whether this could be observable with future space missions dedicated to weak-lensing studies. 

\section{Conclusion}

In this contribution we developed a grid-based potential solver for the quasi-linear formulation of MOND, in order to explore how the potentials of galaxies in the Local Volume out to $\sim 10$~Mpc in Milgromian dynamics are affected by other mass concentrations in the nearby Universe. We solve for the PDM halo in MOND, which in this theory is a convenient abstraction to allow one to readily derive accelerations by using the Newtonian formula. The corresponding `phantom' density (PDM) distribution then appears similar to a $\Lambda$CDM dark matter halo, although with the striking difference that it may also display regions of negative density. 

The concept of PDM is particularly useful as it allows to re-express some predictions of MOND in a Newtonian framework, in terms of relations between the baryonic components of the galaxy and the (phantom) halo mass. We show that MOND predicts a monotonically increasing power-law for the stellar-to-(phantom)halo mass relation (SHMR) and a flat gas-to-halo mass relation. 

We show that the so-called External Field Effect (EFE) of MOND can give rise to PDM densities with surprisingly complex spatial distributions in some situations. However, we find that the dominance of the Virgo supercluster tends to wash out most of these complexities in the Local Volume, generally giving rise to donut-shaped regions of negative phantom density aligned with the direction towards that mass concentration. We also computed the average density of PDM residing around galaxies in the Local Volume $\Omega_{\rm pdm} \approx 0.1$, below the average cold dark matter (CDM) density in a $\Lambda$CDM Universe but comparable to estimates of the DM mass associated with galaxies in the standard context.

In the Milky-Way Andromeda binary system, we find that a region of negative PDM should be located between the two galaxies, presumably affecting the shape of the outskirts of their stellar halos or hot gas coronae. We also found that the LMC PDM mass prior to its current infall was of the order of $8 \times 10^{10} {\rm M}_\odot$, which should be sufficient to affect the relax motion of the MW and induce a signature in the stellar halo. However, the LMC is ``losing" PDM when it penetrates in the MW halo, so that the signature would probably be weaker than in the standard context. We leave to further work a detailed modelling of the effect of the LMC on the MW dynamics in the MOND context.

Finally, we show that probably the most promising way to detect the negative PDM densities predicted by Milgromian gravity would be through weak-lensing. By placing a galaxy of baryonic mass $5\times10^{10}$ M$_\odot$ under an EFE of $g_{Ne}=0.02$ $a_0$, the gravitational lensing convergence parameter $\kappa$ can reach negative values of the order of $-3\times10^{-3}$.

We conclude that if the Local Universe is Milgromian, looking at it with Newtonian eyes can give a broad picture which is surprisingly similar to the standard one. There are nevertheless some noticeable differences. We for instance make for the first time the prediction that the gas-to-(phantom)halo mass relation expected in MOND should be roughly flat. This however, would not necessarily be impossible to explain in the standard context. We highlight that concave weak-lensing with negative convergence at specific locations correlated with the large-scale environment would on the other hand be a true smoking gun of MOND, which might be achievable with future weak-lensing surveys.

\section*{Acknowledgments}

The authors acknowledge enlightening discussions with HongSheng Zhao, and thank the referee for comments and suggestions which helped improve the manuscript. PAO, BF, RI, JF, LP and GM acknowledge funding from the European Research Council (ERC) under the European Unions Horizon 2020 research and innovation programme (grant agreement No. 834148) and from the Agence Nationale de la Recherche (ANR projects ANR-18-CE31-0006 and ANR-19-CE31-0017). This work has received support from the Partenariat Hubert Curien (PHC) for PROCOPE project 44677UE and the Deutscher Akademischer Austauschdienst for PPP grant 57512596 funded by the Bundesministerium fur Bildung und Forschung. GFT acknowledges support from the Agencia Estatal de Investigaci\'on (AEI) of the Ministerio de Ciencia e Innovaci\'on (MCINN) under grant with reference (FJC2018-037323-I). MK acknowledges funding from the Schlumberger Foundation Faculty for the Future program. OM is grateful to the Swiss National Science Foundation for financial support. NIL acknowledges financial support of the Project IDEXLYON at the University of Lyon under the Investments for the Future Program (ANR-16-IDEX-0005). M.S.P. acknowledges funding of a Leibniz-Junior Research Group via the Leibniz Competition, and thanks the Klaus Tschira Stiftung and German Scholars Organization for support via a Klaus Tschira Boost Fund. The authors would also like to acknowledge the High Performance Computing Center of the University of Strasbourg for supporting this work by providing scientific support and access to computing resources. Part of the computing resources were funded by the Equipex Equip@Meso project (Programme Investissements d'Avenir) and the CPER Alsacalcul/Big Data. 


\bibliographystyle{aasjournal}
\bibliography{ms}

\begin{thebibliography}{}
\expandafter\ifx\csname natexlab\endcsname\relax\def\natexlab#1{#1}\fi
\providecommand{\url}[1]{\href{#1}{#1}}

\bibitem[{{Angus} {et~al.}(2008){Angus}, {Famaey}, \&
  {Buote}}]{2008MNRAS.387.1470A}
{Angus}, G.~W., {Famaey}, B., \& {Buote}, D.~A. 2008, \mnras, 387, 1470

\bibitem[{{Angus} {et~al.}(2010){Angus}, {Famaey}, \&
  {Diaferio}}]{2010MNRAS.402..395A}
{Angus}, G.~W., {Famaey}, B., \& {Diaferio}, A. 2010, \mnras, 402, 395

\bibitem[{{Angus} {et~al.}(2012){Angus}, {van der Heyden}, {Famaey}, {Gentile},
  {McGaugh}, \& {de Blok}}]{2012MNRAS.421.2598A}
{Angus}, G.~W., {van der Heyden}, K.~J., {Famaey}, B., {et~al.} 2012, \mnras,
  421, 2598

\bibitem[{{Banik} {et~al.}(2020){Banik}, {Thies}, {Famaey}, {Candlish},
  {Kroupa}, \& {Ibata}}]{2020ApJ...905..135B}
{Banik}, I., {Thies}, I., {Famaey}, B., {et~al.} 2020, \apj, 905, 135

\bibitem[{{Banik} \& {Zhao}(2018{\natexlab{a}})}]{2018MNRAS.473..419B}
{Banik}, I., \& {Zhao}, H. 2018{\natexlab{a}}, \mnras, 473, 419

\bibitem[{{Banik} \& {Zhao}(2018{\natexlab{b}})}]{2018MNRAS.480.2660B}
---. 2018{\natexlab{b}}, \mnras, 480, 2660

\bibitem[{{Behroozi} {et~al.}(2013){Behroozi}, {Wechsler}, \&
  {Conroy}}]{2013ApJ...770...57B}
{Behroozi}, P.~S., {Wechsler}, R.~H., \& {Conroy}, C. 2013, \apj, 770, 57

\bibitem[{{Bekenstein} \& {Milgrom}(1984)}]{1984ApJ...286....7B}
{Bekenstein}, J., \& {Milgrom}, M. 1984, \apj, 286, 7

\bibitem[{{B{\'\i}lek} {et~al.}(2019){B{\'\i}lek}, {M{\"u}ller}, \&
  {Famaey}}]{2019A&A...627L...1B}
{B{\'\i}lek}, M., {M{\"u}ller}, O., \& {Famaey}, B. 2019, \aap, 627, L1

\bibitem[{{Binney} \& {Tremaine}(2008)}]{2008gady.book.....B}
{Binney}, J., \& {Tremaine}, S. 2008, {Galactic Dynamics: Second Edition}

\bibitem[{{Bowman} {et~al.}(2018){Bowman}, {Rogers}, {Monsalve}, {Mozdzen}, \&
  {Mahesh}}]{2018Natur.555...67B}
{Bowman}, J.~D., {Rogers}, A. E.~E., {Monsalve}, R.~A., {Mozdzen}, T.~J., \&
  {Mahesh}, N. 2018, \nat, 555, 67

\bibitem[{{Brada} \& {Milgrom}(1999)}]{1999ApJ...519..590B}
{Brada}, R., \& {Milgrom}, M. 1999, \apj, 519, 590

\bibitem[{{Brada} \& {Milgrom}(2000)}]{2000ApJ...531L..21B}
---. 2000, \apjl, 531, L21

\bibitem[{{Bullock} \& {Boylan-Kolchin}(2017)}]{2017ARA&A..55..343B}
{Bullock}, J.~S., \& {Boylan-Kolchin}, M. 2017, \araa, 55, 343

\bibitem[{{Candlish} {et~al.}(2015){Candlish}, {Smith}, \&
  {Fellhauer}}]{2015MNRAS.446.1060C}
{Candlish}, G.~N., {Smith}, R., \& {Fellhauer}, M. 2015, \mnras, 446, 1060

\bibitem[{{Chae} {et~al.}(2021){Chae}, {Desmond}, {Lelli}, {McGaugh}, \&
  {Schombert}}]{2021arXiv210904745C}
{Chae}, K.-H., {Desmond}, H., {Lelli}, F., {McGaugh}, S.~S., \& {Schombert},
  J.~M. 2021, arXiv e-prints, arXiv:2109.04745

\bibitem[{{Chae} {et~al.}(2020){Chae}, {Lelli}, {Desmond}, {McGaugh}, {Li}, \&
  {Schombert}}]{2020ApJ...904...51C}
{Chae}, K.-H., {Lelli}, F., {Desmond}, H., {et~al.} 2020, \apj, 904, 51

\bibitem[{{Collins} {et~al.}(2014){Collins}, {Chapman}, {Rich}, {Ibata},
  {Martin}, {Irwin}, {Bate}, {Lewis}, {Pe{\~n}arrubia}, {Arimoto}, {Casey},
  {Ferguson}, {Koch}, {McConnachie}, \& {Tanvir}}]{2014ApJ...783....7C}
{Collins}, M. L.~M., {Chapman}, S.~C., {Rich}, R.~M., {et~al.} 2014, \apj, 783,
  7

\bibitem[{{Combes}(2014)}]{2014A&A...571A..82C}
{Combes}, F. 2014, \aap, 571, A82

\bibitem[{{Combes}(2016)}]{2016ASSL..418..413C}
---. 2016, {Explaining the Formation of Bulges with MOND}, ed.
  E.~{Laurikainen}, R.~{Peletier}, \& D.~{Gadotti}, Vol. 418, 413

\bibitem[{{Desmond}(2017{\natexlab{a}})}]{2017MNRAS.472L..35D}
{Desmond}, H. 2017{\natexlab{a}}, \mnras, 472, L35

\bibitem[{{Desmond}(2017{\natexlab{b}})}]{2017MNRAS.464.4160D}
---. 2017{\natexlab{b}}, \mnras, 464, 4160

\bibitem[{{Desmond} {et~al.}(2018){Desmond}, {Ferreira}, {Lavaux}, \&
  {Jasche}}]{2018MNRAS.474.3152D}
{Desmond}, H., {Ferreira}, P.~G., {Lavaux}, G., \& {Jasche}, J. 2018, \mnras,
  474, 3152

\bibitem[{{Di Cintio} \& {Lelli}(2016)}]{2016MNRAS.456L.127D}
{Di Cintio}, A., \& {Lelli}, F. 2016, \mnras, 456, L127

\bibitem[{{Famaey} \& {Binney}(2005)}]{2005MNRAS.363..603F}
{Famaey}, B., \& {Binney}, J. 2005, \mnras, 363, 603

\bibitem[{{Famaey} {et~al.}(2007){Famaey}, {Bruneton}, \&
  {Zhao}}]{2007MNRAS.377L..79F}
{Famaey}, B., {Bruneton}, J.-P., \& {Zhao}, H. 2007, \mnras, 377, L79

\bibitem[{{Famaey} {et~al.}(2018){Famaey}, {McGaugh}, \&
  {Milgrom}}]{2018MNRAS.480..473F}
{Famaey}, B., {McGaugh}, S., \& {Milgrom}, M. 2018, \mnras, 480, 473

\bibitem[{{Famaey} \& {McGaugh}(2012)}]{2012LRR....15...10F}
{Famaey}, B., \& {McGaugh}, S.~S. 2012, Living Reviews in Relativity, 15, 10

\bibitem[{{Freundlich} {et~al.}(2021){Freundlich}, {Famaey}, {Oria},
  {B{\'\i}lek}, {M{\"u}ller}, \& {Ibata}}]{2021arXiv210904487F}
{Freundlich}, J., {Famaey}, B., {Oria}, P.~A., {et~al.} 2021, arXiv e-prints,
  arXiv:2109.04487

\bibitem[{{Fukugita} \& {Peebles}(2004)}]{2004ApJ...616..643F}
{Fukugita}, M., \& {Peebles}, P.~J.~E. 2004, \apj, 616, 643

\bibitem[{{Garavito-Camargo} {et~al.}(2019){Garavito-Camargo}, {Besla},
  {Laporte}, {Johnston}, {G{\'o}mez}, \& {Watkins}}]{2019ApJ...884...51G}
{Garavito-Camargo}, N., {Besla}, G., {Laporte}, C. F.~P., {et~al.} 2019, \apj,
  884, 51

\bibitem[{{Gentile} {et~al.}(2011){Gentile}, {Famaey}, \& {de
  Blok}}]{2011A&A...527A..76G}
{Gentile}, G., {Famaey}, B., \& {de Blok}, W.~J.~G. 2011, \aap, 527, A76

\bibitem[{{Ghari} {et~al.}(2019){Ghari}, {Famaey}, {Laporte}, \&
  {Haghi}}]{2019A&A...623A.123G}
{Ghari}, A., {Famaey}, B., {Laporte}, C., \& {Haghi}, H. 2019, \aap, 623, A123

\bibitem[{{Haslbauer} {et~al.}(2020){Haslbauer}, {Banik}, \&
  {Kroupa}}]{2020MNRAS.499.2845H}
{Haslbauer}, M., {Banik}, I., \& {Kroupa}, P. 2020, \mnras, 499, 2845

\bibitem[{{Hees} {et~al.}(2016){Hees}, {Famaey}, {Angus}, \&
  {Gentile}}]{2016MNRAS.455..449H}
{Hees}, A., {Famaey}, B., {Angus}, G.~W., \& {Gentile}, G. 2016, \mnras, 455,
  449

\bibitem[{{Hoekstra}(2013)}]{2013arXiv1312.5981H}
{Hoekstra}, H. 2013, arXiv e-prints, arXiv:1312.5981

\bibitem[{{Huang} {et~al.}(2016){Huang}, {Liu}, {Yuan}, {Xiang}, {Zhang},
  {Chen}, {Ren}, {Wang}, {Zhang}, {Hou}, {Wang}, \&
  {Cao}}]{2016MNRAS.463.2623H}
{Huang}, Y., {Liu}, X.~W., {Yuan}, H.~B., {et~al.} 2016, \mnras, 463, 2623

\bibitem[{{Irrgang} {et~al.}(2013){Irrgang}, {Wilcox}, {Tucker}, \&
  {Schiefelbein}}]{2013A&A...549A.137I}
{Irrgang}, A., {Wilcox}, B., {Tucker}, E., \& {Schiefelbein}, L. 2013, \aap,
  549, A137

\bibitem[{{Izumi} {et~al.}(2013){Izumi}, {Hagiwara}, {Nakajima}, {Kitamura}, \&
  {Asada}}]{2013PhRvD..88b4049I}
{Izumi}, K., {Hagiwara}, C., {Nakajima}, K., {Kitamura}, T., \& {Asada}, H.
  2013, \prd, 88, 024049

\bibitem[{{Karachentsev} {et~al.}(2013){Karachentsev}, {Makarov}, \&
  {Kaisina}}]{2013AJ....145..101K}
{Karachentsev}, I.~D., {Makarov}, D.~I., \& {Kaisina}, E.~I. 2013, \aj, 145,
  101

\bibitem[{{Karachentsev} \& {Telikova}(2018)}]{2018AN....339..615K}
{Karachentsev}, I.~D., \& {Telikova}, K.~N. 2018, Astronomische Nachrichten,
  339, 615

\bibitem[{{Keller} \& {Wadsley}(2017)}]{2017ApJ...835L..17K}
{Keller}, B.~W., \& {Wadsley}, J.~W. 2017, \apjl, 835, L17

\bibitem[{{Kravtsov} {et~al.}(2004){Kravtsov}, {Gnedin}, \&
  {Klypin}}]{2004ApJ...609..482K}
{Kravtsov}, A.~V., {Gnedin}, O.~Y., \& {Klypin}, A.~A. 2004, \apj, 609, 482

\bibitem[{{Lelli} {et~al.}(2016{\natexlab{a}}){Lelli}, {McGaugh}, \&
  {Schombert}}]{2016ApJ...816L..14L}
{Lelli}, F., {McGaugh}, S.~S., \& {Schombert}, J.~M. 2016{\natexlab{a}}, \apjl,
  816, L14

\bibitem[{{Lelli} {et~al.}(2016{\natexlab{b}}){Lelli}, {McGaugh}, \&
  {Schombert}}]{2016AJ....152..157L}
---. 2016{\natexlab{b}}, \aj, 152, 157

\bibitem[{{Lelli} {et~al.}(2019){Lelli}, {McGaugh}, {Schombert}, {Desmond}, \&
  {Katz}}]{2019MNRAS.484.3267L}
{Lelli}, F., {McGaugh}, S.~S., {Schombert}, J.~M., {Desmond}, H., \& {Katz}, H.
  2019, \mnras, 484, 3267

\bibitem[{{Lelli} {et~al.}(2016{\natexlab{c}}){Lelli}, {McGaugh}, {Schombert},
  \& {Pawlowski}}]{2016ApJ...827L..19L}
{Lelli}, F., {McGaugh}, S.~S., {Schombert}, J.~M., \& {Pawlowski}, M.~S.
  2016{\natexlab{c}}, \apjl, 827, L19

\bibitem[{{Lelli} {et~al.}(2017){Lelli}, {McGaugh}, {Schombert}, \&
  {Pawlowski}}]{2017ApJ...836..152L}
---. 2017, \apj, 836, 152

\bibitem[{{Li} {et~al.}(2018){Li}, {Lelli}, {McGaugh}, \&
  {Schombert}}]{2018A&A...615A...3L}
{Li}, P., {Lelli}, F., {McGaugh}, S., \& {Schombert}, J. 2018, \aap, 615, A3

\bibitem[{{Llinares} {et~al.}(2008){Llinares}, {Knebe}, \&
  {Zhao}}]{2008MNRAS.391.1778L}
{Llinares}, C., {Knebe}, A., \& {Zhao}, H. 2008, \mnras, 391, 1778

\bibitem[{{Londrillo} \& {Nipoti}(2009)}]{2009MSAIS..13...89L}
{Londrillo}, P., \& {Nipoti}, C. 2009, Memorie della Societa Astronomica
  Italiana Supplementi, 13, 89

\bibitem[{{Ludlow} {et~al.}(2017){Ludlow}, {Ben{\'\i}tez-Llambay}, {Schaller},
  {Theuns}, {Frenk}, {Bower}, {Schaye}, {Crain}, {Navarro}, {Fattahi}, \&
  {Oman}}]{2017PhRvL.118p1103L}
{Ludlow}, A.~D., {Ben{\'\i}tez-Llambay}, A., {Schaller}, M., {et~al.} 2017,
  \prl, 118, 161103

\bibitem[{{L{\"u}ghausen} {et~al.}(2015){L{\"u}ghausen}, {Famaey}, \&
  {Kroupa}}]{2015CaJPh..93..232L}
{L{\"u}ghausen}, F., {Famaey}, B., \& {Kroupa}, P. 2015, Canadian Journal of
  Physics, 93, 232

\bibitem[{{Marasco} {et~al.}(2020){Marasco}, {Posti}, {Oman}, {Famaey},
  {Cresci}, \& {Fraternali}}]{2020A&A...640A..70M}
{Marasco}, A., {Posti}, L., {Oman}, K., {et~al.} 2020, \aap, 640, A70

\bibitem[{{McGaugh} \& {Milgrom}(2013{\natexlab{a}})}]{2013ApJ...766...22M}
{McGaugh}, S., \& {Milgrom}, M. 2013{\natexlab{a}}, \apj, 766, 22

\bibitem[{{McGaugh} \& {Milgrom}(2013{\natexlab{b}})}]{2013ApJ...775..139M}
---. 2013{\natexlab{b}}, \apj, 775, 139

\bibitem[{{McGaugh} {et~al.}(2016){McGaugh}, {Lelli}, \&
  {Schombert}}]{2016PhRvL.117t1101M}
{McGaugh}, S.~S., {Lelli}, F., \& {Schombert}, J.~M. 2016, \prl, 117, 201101

\bibitem[{{McGaugh} {et~al.}(2000){McGaugh}, {Schombert}, {Bothun}, \& {de
  Blok}}]{2000ApJ...533L..99M}
{McGaugh}, S.~S., {Schombert}, J.~M., {Bothun}, G.~D., \& {de Blok}, W.~J.~G.
  2000, \apjl, 533, L99

\bibitem[{{McGaugh} \& {van Dokkum}(2021)}]{2021RNAAS...5...23M}
{McGaugh}, S.~S., \& {van Dokkum}, P. 2021, Research Notes of the American
  Astronomical Society, 5, 23

\bibitem[{{Milgrom}(1983)}]{1983ApJ...270..365M}
{Milgrom}, M. 1983, \apj, 270, 365

\bibitem[{{Milgrom}(1986)}]{1986ApJ...306....9M}
---. 1986, \apj, 306, 9

\bibitem[{{Milgrom}(2008)}]{2008NewAR..51..906M}
---. 2008, \nar, 51, 906

\bibitem[{{Milgrom}(2010)}]{2010MNRAS.403..886M}
---. 2010, \mnras, 403, 886

\bibitem[{{Milgrom}(2016)}]{2016PhRvL.117n1101M}
---. 2016, \prl, 117, 141101

\bibitem[{{Miyamoto} \& {Nagai}(1975)}]{1975PASJ...27..533M}
{Miyamoto}, M., \& {Nagai}, R. 1975, \pasj, 27, 533

\bibitem[{{Monari} {et~al.}(2018){Monari}, {Famaey}, {Carrillo}, {Piffl},
  {Steinmetz}, {Wyse}, {Anders}, {Chiappini}, \&
  {Jan{\ss}en}}]{2018A&A...616L...9M}
{Monari}, G., {Famaey}, B., {Carrillo}, I., {et~al.} 2018, \aap, 616, L9

\bibitem[{{Moster} {et~al.}(2013){Moster}, {Naab}, \&
  {White}}]{2013MNRAS.428.3121M}
{Moster}, B.~P., {Naab}, T., \& {White}, S. D.~M. 2013, \mnras, 428, 3121

\bibitem[{{Nakajima} {et~al.}(2014){Nakajima}, {Izumi}, \&
  {Asada}}]{2014PhRvD..90h4026N}
{Nakajima}, K., {Izumi}, K., \& {Asada}, H. 2014, \prd, 90, 084026

\bibitem[{{Navarro} {et~al.}(2017){Navarro}, {Ben{\'\i}tez-Llambay}, {Fattahi},
  {Frenk}, {Ludlow}, {Oman}, {Schaller}, \& {Theuns}}]{2017MNRAS.471.1841N}
{Navarro}, J.~F., {Ben{\'\i}tez-Llambay}, A., {Fattahi}, A., {et~al.} 2017,
  \mnras, 471, 1841

\bibitem[{{Navarro} {et~al.}(1997){Navarro}, {Frenk}, \&
  {White}}]{1997ApJ...490..493N}
{Navarro}, J.~F., {Frenk}, C.~S., \& {White}, S. D.~M. 1997, \apj, 490, 493

\bibitem[{{Oman} {et~al.}(2015){Oman}, {Navarro}, {Fattahi}, {Frenk}, {Sawala},
  {White}, {Bower}, {Crain}, {Furlong}, {Schaller}, {Schaye}, \&
  {Theuns}}]{2015MNRAS.452.3650O}
{Oman}, K.~A., {Navarro}, J.~F., {Fattahi}, A., {et~al.} 2015, \mnras, 452,
  3650

\bibitem[{{Papastergis} {et~al.}(2012){Papastergis}, {Cattaneo}, {Huang},
  {Giovanelli}, \& {Haynes}}]{2012ApJ...759..138P}
{Papastergis}, E., {Cattaneo}, A., {Huang}, S., {Giovanelli}, R., \& {Haynes},
  M.~P. 2012, \apj, 759, 138

\bibitem[{{Pawlowski}(2018)}]{2018MPLA...3330004P}
{Pawlowski}, M.~S. 2018, Modern Physics Letters A, 33, 1830004

\bibitem[{{Peebles}(2020)}]{2020MNRAS.498.4386P}
{Peebles}, P.~J.~E. 2020, \mnras, 498, 4386

\bibitem[{{Piffaretti} {et~al.}(2011){Piffaretti}, {Arnaud}, {Pratt},
  {Pointecouteau}, \& {Melin}}]{2011A&A...534A.109P}
{Piffaretti}, R., {Arnaud}, M., {Pratt}, G.~W., {Pointecouteau}, E., \&
  {Melin}, J.~B. 2011, \aap, 534, A109

\bibitem[{{Posti} \& {Fall}(2021)}]{2021A&A...649A.119P}
{Posti}, L., \& {Fall}, S.~M. 2021, \aap, 649, A119

\bibitem[{{Posti} {et~al.}(2019{\natexlab{a}}){Posti}, {Fraternali}, \&
  {Marasco}}]{2019A&A...626A..56P}
{Posti}, L., {Fraternali}, F., \& {Marasco}, A. 2019{\natexlab{a}}, \aap, 626,
  A56

\bibitem[{{Posti} {et~al.}(2019{\natexlab{b}}){Posti}, {Marasco}, {Fraternali},
  \& {Famaey}}]{2019A&A...629A..59P}
{Posti}, L., {Marasco}, A., {Fraternali}, F., \& {Famaey}, B.
  2019{\natexlab{b}}, \aap, 629, A59

\bibitem[{{Ramachandran} \& {Varoquaux}(2011)}]{2011CSE....13b..40R}
{Ramachandran}, P., \& {Varoquaux}, G. 2011, Computing in Science and
  Engineering, 13, 40

\bibitem[{{Riess} {et~al.}(2019){Riess}, {Casertano}, {Yuan}, {Macri}, \&
  {Scolnic}}]{2019ApJ...876...85R}
{Riess}, A.~G., {Casertano}, S., {Yuan}, W., {Macri}, L.~M., \& {Scolnic}, D.
  2019, \apj, 876, 85

\bibitem[{{Sanders}(1999)}]{1999ApJ...512L..23S}
{Sanders}, R.~H. 1999, \apjl, 512, L23

\bibitem[{{Skordis} \& {Zlosnik}(2020)}]{2020arXiv200700082S}
{Skordis}, C., \& {Zlosnik}, T. 2020, arXiv e-prints, arXiv:2007.00082

\bibitem[{{Thomas} {et~al.}(2017){Thomas}, {Famaey}, {Ibata}, {L{\"u}ghausen},
  \& {Kroupa}}]{2017A&A...603A..65T}
{Thomas}, G.~F., {Famaey}, B., {Ibata}, R., {L{\"u}ghausen}, F., \& {Kroupa},
  P. 2017, \aap, 603, A65

\bibitem[{{Thomas} {et~al.}(2018){Thomas}, {Famaey}, {Ibata}, {Renaud},
  {Martin}, \& {Kroupa}}]{2018A&A...609A..44T}
{Thomas}, G.~F., {Famaey}, B., {Ibata}, R., {et~al.} 2018, \aap, 609, A44

\bibitem[{{Tiret} \& {Combes}(2007)}]{2007A&A...464..517T}
{Tiret}, O., \& {Combes}, F. 2007, \aap, 464, 517

\bibitem[{{Tully} {et~al.}(2016){Tully}, {Courtois}, \&
  {Sorce}}]{2016AJ....152...50T}
{Tully}, R.~B., {Courtois}, H.~M., \& {Sorce}, J.~G. 2016, \aj, 152, 50

\bibitem[{{Vale} \& {Ostriker}(2004)}]{2004MNRAS.353..189V}
{Vale}, A., \& {Ostriker}, J.~P. 2004, \mnras, 353, 189

\bibitem[{{Willmer}(2018)}]{2018ApJS..236...47W}
{Willmer}, C. N.~A. 2018, \apjs, 236, 47

\bibitem[{{Wu} {et~al.}(2008){Wu}, {Famaey}, {Gentile}, {Perets}, \&
  {Zhao}}]{2008MNRAS.386.2199W}
{Wu}, X., {Famaey}, B., {Gentile}, G., {Perets}, H., \& {Zhao}, H. 2008,
  \mnras, 386, 2199

\bibitem[{{Wu} \& {Kroupa}(2015)}]{2015MNRAS.446..330W}
{Wu}, X., \& {Kroupa}, P. 2015, \mnras, 446, 330

\bibitem[{{Wu} {et~al.}(2017){Wu}, {Wang}, {Feix}, \&
  {Zhao}}]{2017ApJ...844..130W}
{Wu}, X., {Wang}, Y., {Feix}, M., \& {Zhao}, H. 2017, \apj, 844, 130

\bibitem[{{Xue} {et~al.}(2008){Xue}, {Rix}, {Zhao}, {Re Fiorentin}, {Naab},
  {Steinmetz}, {van den Bosch}, {Beers}, {Lee}, {Bell}, {Rockosi}, {Yanny},
  {Newberg}, {Wilhelm}, {Kang}, {Smith}, \& {Schneider}}]{2008ApJ...684.1143X}
{Xue}, X.~X., {Rix}, H.~W., {Zhao}, G., {et~al.} 2008, \apj, 684, 1143

\bibitem[{{Zhao} \& {Famaey}(2010)}]{2010PhRvD..81h7304Z}
{Zhao}, H., \& {Famaey}, B. 2010, \prd, 81, 087304

\end{thebibliography}

\appendix

\section{List of UNGC sources}

\begin{longtable}{l|c|c|>{\centering\arraybackslash}p{2cm}|>{\centering\arraybackslash}p{2cm}|c|>{\centering\arraybackslash}p{2cm}}%
\caption{List of sources extracted from \citet{2013AJ....145..101K} and processed as explained in Subsection~\ref{subsec:UNGC}. Columns are galaxy name, distance to MW, baryonic mass, PDM mass computed numerically up to $r_{200}$ with EFE, associated $r_{200}$, Newtonian external gravitational acceleration $g_{Ne}$ in units of $a_0$ at galaxy's location, and PDM mass computed numerically up to $r_{200}$ without EFE. Bottom entries with N/A are galaxies too embedded in another galaxy's PDM halo for it to make sense to compute their PDM mass.}
\label{table:local_sources}\\
    Galaxy & d (kpc) & $M_b$ ($M_{\odot}$) & $M_{\rm PDM}$ ($M_{\odot}$) (with EFE) & $r_{200}$ (kpc) (with EFE) & $g_{Ne}/a_0$ & $M_{\rm PDM}$ ($M_{\odot}$) (no EFE) 
    \csvreader[head to column names]{local_sources.csv}{} 
    {\\\hline\csvcoli & \csvcolii & \csvcoliii & \csvcoliv & \csvcolv & \csvcolvi & \csvcolvii}
\end{longtable}

\section{List of galaxy clusters}

\begin{longtable}{l|c|c|c|c|c}
\caption{List of sources outside of the Local Volume. For galaxy clusters, the given galaxy is the one chosen for the cluster center. Columns are object name (cluster name), distance to MW, MOND mass, $r_{200}$, $g_N/a_0$ at the center of the MW, $g_N/a_0$ at the cluster's $r_{200}$.}
\label{table:external_sources}\\
    Object & d (kpc) & $M_{\rm MOND}$ ($M_{\odot}$) & $r_{200}$ (kpc) & $g_N/a_0$ at MW & $g_N/a_0$ at $r_{200}$
    \csvreader[head to column names]{external_sources.csv}{} 
    {\\\hline\csvcoli & \csvcolv & \csvcolviii & \csvcolvii & \csvcolx & \csvcolxi}
\end{longtable}

\section{The local universe PDM map with exclusively baryonic galaxy clusters}

\begin{figure*}[!htb]
\centering
  \includegraphics[angle=0,  clip, scale = 0.40]{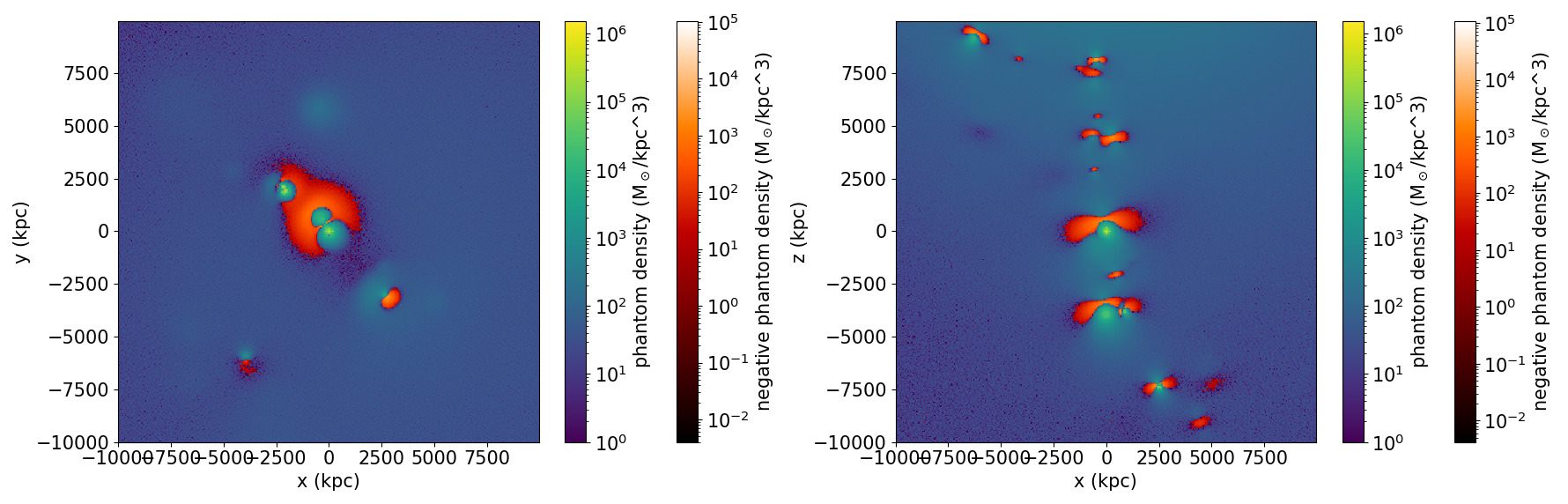}
   \caption{Phantom dark matter density map of the Local Volume centred on the MW (0,0,0). Plane cuts. \textit{Left panel:} MW galactic plane z=0. \textit{Right panel:} MW edge-on view y=0. In this case, clusters are taken into account with their baryonic mass only. This leads to an intermediate situation between those of Figure~\ref{fig:10Mpc_no_ext_sources} and Figure~\ref{fig:10Mpc_everything}, in which the most massive galaxies in the Local Universe regain importance compared to the case where the clusters have their MONDian mass, but the clusters still remain the most important sources of EFE.}
\label{fig:ext_sources_baryonic_mass}
\end{figure*}

\section{Shear map in the isolated MOND case}
\begin{figure*}[!htb]
\centering
  \includegraphics[angle=0,  clip, width=15cm]{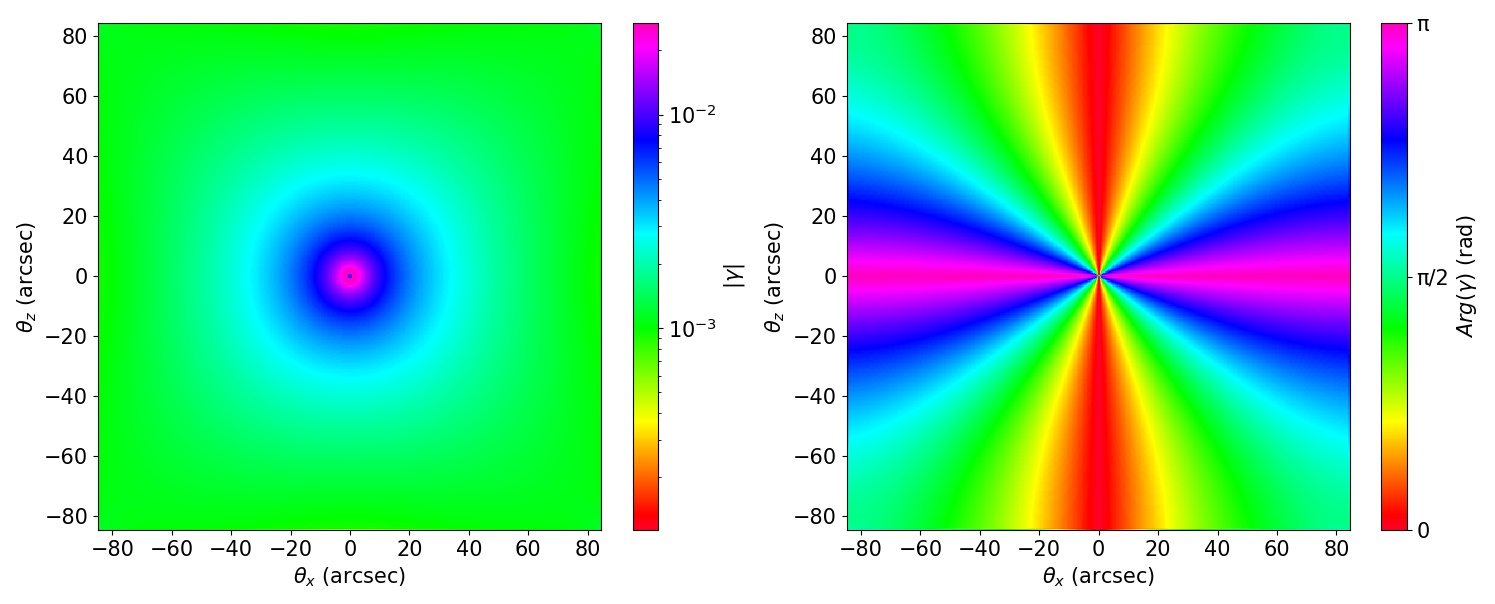}
   \caption{Shear vector $\gamma$ for an NGC5055-like galaxy centred at $\theta_x=\theta_z=0$ in the isolated MOND case (i.e. no EFE), used as a gravitational lens at $z = 0.3$ for sources at $z = 5$. \textit{Left panel:} shear amplitude $|\gamma|$. \textit{Right panel:} shear angle ${\rm Arg}(\gamma)$.}
\label{fig:shear_isolated}
\end{figure*}

As a comparison to Fig.~\ref{fig:shear_EFE}, we provide the shear map corresponding to the same galaxy in the isolated MOND case in Fig.~\ref{fig:shear_isolated}.

\end{document}